%% file: paper.tex
\documentclass[a4paper,journal]{IEEEtran}

\makeatletter
\newcommand{\AddInputPath}[1]{%
  \ifx\input@path\@undefined
    \def\input@path{#1}
  \else
    \g@addto@macro{\input@path}{#1}
  \fi
}
\makeatother

\usepackage{etex}

\usepackage[font=footnotesize,caption=false]{subfig} 

\usepackage{relsize}

\usepackage{array}
\usepackage{booktabs,tabularx}
\usepackage{multirow}
\usepackage{units}
\usepackage{graphicx}

\usepackage[T1]{fontenc}
\usepackage{textcomp}
\usepackage[utf8]{inputenc}
\usepackage[final]{microtype}
\usepackage{icomma}
\usepackage{xspace}

\usepackage[tbtags]{amsmath}
\usepackage{amssymb,amsfonts,bm}
\usepackage{mathtools} 
\usepackage{dsfont}
\usepackage{mathrsfs}
\usepackage{accents}
\usepackage{empheq}
\usepackage{nccmath}
\usepackage{balance}

\usepackage{color}
\usepackage{calc}
\usepackage{tikz}
\usepackage{pgfplots,pgfplotstable}

\usepackage[capitalize]{cleveref}
\usepackage{refcount}

\usepackage[inline,shortlabels]{enumitem}
\usepackage{algorithm}
\usepackage{algpseudocode}
\makeatletter
\newcommand{\algmargin}{\the\ALG@thistlm}
\makeatother
\newlength{\whilewidth}
\algdef{SE}[parWHILE]{parWhile}{EndparWhile}[1]
  {\parbox[t]{\dimexpr\linewidth-\algmargin}{%
     \hangindent\whilewidth\strut\algorithmicwhile\ #1\ \algorithmicdo\strut}}{\algorithmicend\ \algorithmicwhile}%
\algnewcommand{\parState}[1]{\State%
  \parbox[t]{\dimexpr\linewidth-\algmargin}{\strut #1\strut}}

\usepackage{glossaries}
\usepackage{ifthen}
\usepackage{cite}
\usepackage{multibib}
\usepackage[shortcuts]{extdash}

\usepackage{comment}
\usepackage{todonotes}
\let\legacytodo\todo
\newcommand{\ruggedtodo}[2][]{\tikzexternaldisable\legacytodo[#1]{#2}\tikzexternalenable}
\renewcommand{\todo}[1]{\ruggedtodo[inline]{#1}}

\makeatletter
\def\todoref{\@ifnextchar[{\todoref@with}{\todoref@without}}
\def\todoref@without{\textbf{\color{red} [reference needed]}\xspace}
\def\todoref@with[#1]{\textbf{\color{red} [reference needed: #1]}\xspace}
\makeatother

\input{acronyms}
\newacronym{ian}{IAN}{treating interference as noise}
\glsenableentrycount
\makeglossaries

\usetikzlibrary{positioning}
\usetikzlibrary{calc}
\usetikzlibrary{fit}
\usetikzlibrary{intersections}
\usetikzlibrary{decorations.pathreplacing}
\usetikzlibrary{pgfplots.fillbetween}
\usetikzlibrary{bending}
\usetikzlibrary{arrows.meta}

\usetikzlibrary{external}

\pgfkeys{/pgfplots/.cd,
	hk/.style={blue,mark=*},
	snd/.style={red,mark=square*},
	ian/.style={brown!60!black,mark=triangle*},
}

\makeatletter
\newcommand\transformxdimension[1]{
    \pgfmathparse{((#1/\pgfplots@x@veclength)+\pgfplots@data@scale@trafo@SHIFT@x)/10^\pgfplots@data@scale@trafo@EXPONENT@x}
}
\newcommand\transformydimension[1]{
    \pgfmathparse{((#1/\pgfplots@y@veclength)+\pgfplots@data@scale@trafo@SHIFT@y)/10^\pgfplots@data@scale@trafo@EXPONENT@y}
}
\makeatother

\crefname{equation}{}{}
\crefrangeformat{equation}{(#3#1#4)--(#5#2#6)}
\crefformat{footnote}{#2\footnotemark[#1]#3}

\newcommand{\abs}[1]{\ensuremath{\left\lvert #1 \right\rvert}}
\newcommand{\norm}[1]{\ensuremath{\left\lVert #1 \right\rVert}}
\DeclareMathOperator\Capa{C}

\DeclareMathOperator\diam{diam}
\DeclareMathOperator\proj{proj}
\DeclareMathOperator\closure{cl}
\DeclareMathOperator\interior{int}
\DeclareMathOperator*{\argmax}{arg\,max}
\DeclareMathOperator*{\argmin}{arg\,min}
\let\card=\abs

\let\vec\bm

\newcommand{\ubar}[1]{\underaccent{\bar}{#1}}

\allowdisplaybreaks[1]

\DeclareFontFamily{U}{mathx}{\hyphenchar\font45}
\DeclareFontShape{U}{mathx}{m}{n}{
      <5> <6> <7> <8> <9> <10>
      <10.95> <12> <14.4> <17.28> <20.74> <24.88>
      mathx10
      }{}
\DeclareSymbolFont{mathx}{U}{mathx}{m}{n}
\DeclareMathSymbol{\bigtimes}{1}{mathx}{"91}

\newtheorem{theorem}{Theorem}
\newtheorem{lemma}{Lemma}

\newtheorem{proposition}{Proposition}

\newtheorem{remark}{Remark}

\newcommand{\xqed}[1]{%
	\leavevmode\unskip\penalty9999 \hbox{}\nobreak\hfill
	\quad\hbox{\ensuremath{#1}}}

\newtheorem{XXXassumption}{Case}
 
\newenvironment{assumption}
	{\begin{XXXassumption}}
	{\xqed{\lozenge}\end{XXXassumption}}

\crefname{XXXassumption}{Case}{Cases}

\newtheorem{XXXexample}{Example}
\newenvironment{example}
	{\begin{XXXexample}}
	{\xqed{\lozenge}\end{XXXexample}}

\newcounter{optimizationproblem}
\newenvironment{optprob}{\begin{equation}\left\{\begin{aligned}}{\end{aligned}\right.\refstepcounter{optimizationproblem}\tag{P\theoptimizationproblem}\end{equation}\ignorespacesafterend}
\newenvironment{optprob*}{\begin{equation*}\left\{\begin{aligned}}{\end{aligned}\right.\end{equation*}\ignorespacesafterend}

\makeatletter
\newcommand{\subalign}[1]{%
  \vcenter{%
    \Let@ \restore@math@cr \default@tag
    \baselineskip\fontdimen10 \scriptfont\tw@
    \advance\baselineskip\fontdimen12 \scriptfont\tw@
    \lineskip\thr@@\fontdimen8 \scriptfont\thr@@
    \lineskiplimit\lineskip
    \ialign{\hfil$\m@th\scriptstyle##$&$\m@th\scriptstyle{}##$\crcr
      #1\crcr
    }%
  }
}
\makeatother

\pgfplotscreateplotcyclelist{default}{%
	blue,mark=*\\%
	red,mark=star\\%
	teal,mark=square*\\%
	brown!60!black,mark=otimes*\\%
}

\newcommand*\qk{_{q(k)}\xspace}%
\newcommand*\lk{_{l(k)}\xspace}%
\newcommand{\AiShort}[3]{\ensuremath{A_{#1}}\xspace}
\newcommand{\BiShort}[3]{\ensuremath{B_{#1}}\xspace}
\newcommand{\CiShort}[3]{\ensuremath{C_{#1}}\xspace}
\newcommand{\DiShort}[3]{\ensuremath{D_{#1}}\xspace}
\newcommand{\Ai}[3]{\ensuremath{\log\!\bigg( 1 + \frac{\abs{h_{#1}}^2 S^p_{#1}}{\gamma_{#1}(\vec S)} \bigg)}\xspace}%
\newcommand{\Bi}[3]{\ensuremath{\log\!\bigg( 1 + \frac{\abs{h_{#1}}^2 (S^p_{#1} + S^c_{#1})}{\gamma_{#1}(\vec S)} \bigg)}\xspace}%
\newcommand{\Ci}[3]{\ensuremath{\log\!\bigg( 1 + \frac{\abs{h_{#1}}^2 S^p_{#1} + \abs{h_{#3}}^2 S^c_{#3}}{\gamma_{#1}(\vec S)} \bigg)}\xspace}%
\newcommand{\Di}[3]{\ensuremath{\log\!\bigg( 1 + \frac{\abs{h_{#1}}^2 (S^p_{#1} + S^c_{#1}) + \abs{h_{#3}}^2 S^c_{#3}}{\gamma_{#1}(\vec S)} \bigg)}\xspace}%

\begin{document}
\title{Efficient Global Optimal Resource Allocation in Non-Orthogonal Interference Networks}

\author{Bho~Matthiesen,~\IEEEmembership{Student Member,~IEEE,}
		and Eduard~A.~Jorswieck,~\IEEEmembership{Senior Member,~IEEE}%
		\thanks{
			Parts of this paper were presented at the 19\textsuperscript{th} IEEE International Workshop on Signal Processing Advances in Wireless Communications, 2018 \cite{spawc18}, and the 2019 IEEE International Conference on Acoustics, Speech and Signal Processing \cite{icassp2019}.
		}%
		\thanks{
			The authors are with the Chair for Communications Theory, Communications Laboratory, Technische Universität Dresden, Dresden, Germany (e-mail: bho.matthiesen@tu-dresden.de, jorswieck@ieee.org).
		}%
		\thanks{
			This work is supported in part by the German Research Foundation (DFG) in the
			Collaborative Research Center 912 ``Highly Adaptive Energy-Efficient Computing,'' and under grant number JO~801/24-1.
		}%
	}

\maketitle

\begin{abstract}
	Many resource allocation tasks are challenging global (i.e., non-convex) optimization problems. The main issue is that the computational complexity of these problems grows exponentially in the number of variables instead of polynomially as for many convex optimization problems. However, often the non-convexity stems only from a subset of variables. Conventional global optimization frameworks like monotonic optimization or DC programming \cite{Tuy2016} treat all variables as global variables and require complicated, problem specific decomposition approaches to exploit the convexity in some variables \cite{camsap17}. To overcome this challenge, we develop an easy-to-use algorithm that inherently differentiates between convex and non-convex variables, preserving the low computational complexity in the number of convex variables. 
	Another issue with these widely used frameworks is that they may suffer from severe numerical problems. We discuss this issue in detail and provide a clear motivating example. The solution to this problem is to replace the traditional approach of finding an $\varepsilon$\=/approximate solution by the novel concept of \emph{$\varepsilon$\=/essential feasibility}. The underlying algorithmic approach is called \gls{sit} algorithm and builds the foundation of our developed algorithm. A further highlight of this algorithm is that it inherently treats fractional objectives making the use of Dinkelbach's iterative algorithm obsolete. Numerical experiments show a speed-up of four orders of magnitude over state-of-the-art algorithms and almost three orders of magnitude of additional speed-up over Dinkelbach's algorithm for fractional programs.
\end{abstract}
\glsresetall
\glsunset{dc}

\begin{IEEEkeywords}
Resource allocation, global optimization, successive incumbent transcending, essential feasibility, multi-way relay channel, simultaneous non-unique decoding, interference networks
\end{IEEEkeywords}

\section{Introduction}
Resource allocation is essential in most communication systems \cite{Han2008}. Practical systems usually use algorithms with no or only weak optimality guarantees for performance reasons. Nevertheless, asserting the quality of these algorithms requires the knowledge of the optimal solution to these problems.
The general optimization problem 
\begin{equation}
	\max_{\vec x\in\mathcal C} f(\vec x)
\refstepcounter{optimizationproblem}\tag{P\theoptimizationproblem}
\label{opt}
\end{equation}
with $\mathcal C\in\mathds R^n$ and $f: \mathds R^n \mapsto \mathds R$ covers a large class of resource allocation problems.
A point $\vec x^\ast\in\mathcal C$ satisfying $f(\vec x^\ast) \ge f(\vec x)$ for all $\vec x\in\mathcal C$ is called a \emph{global maximizer} of $f$. If $\vec x^\ast$ only satisfies this condition for all $\vec x$ in an open $\varepsilon$-neighborhood of $\vec x^\ast$ for some $\varepsilon > 0$, i.e., for all $\vec x \in \{\vec x\in\mathds R^n : \norm{\vec x - \vec x^\ast} < \varepsilon \} \cap \mathcal C$, it is called a \emph{local minimizer}.
The difficulty in obtaining a global optimal solution to \cref{opt} is that all algorithms with polynomial computational complexity can at most obtain a local optimal solution. So, unless \cref{opt} belongs to the class of optimization problems with the property that every local maximum is a global maximum,\footnote{An important example are convex optimization problems where the objective of \cref{opt} is a concave function and $\mathcal C$ is a convex set.} solving \cref{opt} has exponential computational complexity \cite{Horst1996}.

As an example, consider allocating the transmit power in an interference network. Albeit the capacity region of such an network is not known in general, the optimal decoder under the assumption of random codebooks\footnote{The random codebooks are restricted to superposition coding and time sharing.} is known to be \cgls{snd} \cite{bandemer2015}. This leads to a global optimization problem where optimization is done jointly over the rates and powers. A close examination of this problem reveals that it is linear in the rate variables, i.e., for fixed power variables the problem can be solved in polynomial time \cite{Khachiyan1979,Karmarkar1984,camsap17}. Hence, the power variables are the only reason that the optimization problem is global and has exponential complexity.
We call these variables \emph{global} variables, while the remaining ones are named \emph{non-global}.\footnote{A more precise definition is given in \cref{sec:notation}.}
The most popular solution approaches for global resource allocation problems are monotonic optimization and \cgls{dc} programming. Both frameworks treat all variables as global variables which often results in unnecessary high numerical complexity.
Moreover, transforming typical resource allocation problems to fit into these frameworks often requires auxiliary variables,
which, of course, further increases computational complexity.
Instead, in this paper we present a novel framework that preserves the computational complexity of the non-global variables and does not require any auxiliary variables.

Another often neglected issue with these algorithms is the assumption of a robust feasible set, i.e., a set with no isolated points. If this assumption does not hold, which might be the case for resource allocation problems, it leads to serious numerical problems.
We avoid this problem entirely by using robust global optimization\footnote{There are, at least, two different meanings of ``robust optimization:'' the one discussed here that is robust against the effects of non-robust feasible sets and small changes in the tolerances, and the one that provides robustness against uncertainty in the input data \cite{Ben-Tal2009}, e.g., robust beamforming \cite{Vorobyov2003}, or robust monotonic optimization \cite{Bjoernson2012}.} \cite{Tuy2005a,Tuy2009,Tuy2016}. The core idea is to shrink the feasible set by an infinitesimal amount and then solve a sequence of feasibility problems with a \cgls{bb} procedure.
This approach is called \emph{\cgls{sit} scheme} and does not require any assumptions on the robustness of the feasible set $\mathcal C$ because it is designed to operate only on the accumulation points of $\mathcal C$.
The result is a numerically much stabler procedure than could be obtained using classical monotonic or \cgls{dc} programming algorithms. Moreover, the \cgls{sit} approach always provides a good feasible solution even if stopped prematurely. Instead, conventional algorithms usually outer approximate the solution rendering intermediate solutions almost useless (because they are infeasbile).

Fractional objectives, which occur, e.g., in the optimization of the \cgls{ee}, can not be handled directly by monotonic optimization or \cgls{dc} programming. Instead, Dinkelbach's algorithm is used where the original problem is transformed into an auxiliary problem which is then solved several times with one of these frameworks \cite{Zappone2017}. However, this approach has several drawbacks. First, convergence to the optimal solution of the original problem is only guaranteed if the auxiliary problem is solved exactly. In practice, this algorithm also works well for approximate solutions but the numerical accuracy should be sufficiently high. Second, the auxiliary problem needs to be solved several times, and, finally, the stopping criterion is unrelated to the distance of the obtained approximate optimal value to the true optimum. Especially the first two are critical for global optimization since they increase the computation time significantly. Instead, our framework is able to deal directly with fractional objectives avoiding these problems entirely.

\paragraph{Related work}
The \cgls{sit} approach was developed by Hoang Tuy in \cite{Tuy2005a,Tuy2009,Tuy2016} and, to the best of our knowledge, has not been adopted for resource allocation problems yet. However, the importance of robust feasible sets has been noted in \cite{Phan2012} where beamforming in a cognitive radio network is solved with \cgls{dc} programming. In \cite{Xu2010}, the basic principle of the \cgls{sit} approach is used to solve a \cgls{mop}.

Instead, decomposition approaches\cite{Palomar2006} are widely used.
For example, in \cite{Palomar2005} the design of linear transceivers for multicarrier \cgls{mimo} channels is considered. This challenging non-convex problem is solved by primal decomposition into a convex outer problem and inner problems with closed-form solutions.
The authors of \cite{Kaleva2016} combine successive convex approximation and primal decomposition to solve the sum rate maximization problem with \gls{qos} constraints for interfering broadcast channels with first order optimality.
A distributed algorithm for coordinated beamforming in multicell multigroup multicast systems is developed in \cite{Tervo2018} based on primal decomposition and semidefinite relaxation.
In \cite{Rossi2011} the partly convex-monotone structure of utility maximization problems in broadcast and interference channels is exploited via a \cgls{bb} procedure where branching is only performed over the global variables. This approach is similar to our framework but the problem setting is more specific and the optimization is over a convex set, i.e., the \cgls{sit} approach is not needed.
In \cite{camsap17}, we solve a special case of the problems considered here by decomposing it into an inner linear and an outer monotonic program which is solved by the Polyblock algorithm \cite{Tuy2000}.

Resource allocation for interference networks is mostly done under orthogonality assumptions to fall back to the noise-limited case. For example, in \cite{Qian2009} monotonic optimization was first used to maximize the throughput in an interference network where interference is treated as noise. In \cite{Bjornson2013} a monotonic optimization based framework for resource allocation in coordinated multi-cell systems is presented.
Optimization of the \cgls{ee} in interference networks under the assumption that interference is treated as noise is considered in \cite{Zappone2017,Zappone2016}.
In \cite{Tervo2017} energy-efficient coordinated beamforming in multi-cell, multi-user systems is considered under the assumption of realistic power consumption models. Energy-efficient resource allocation in OFDMA systems with and without wireless power transfer is studied in \cite{Ng2013} and \cite{Ng2012}, respectively.
This is, naturally, just a very incomplete list of papers dealing with
resource allocation problems in interference networks under orthogonality constraints.
Of course, the reasons for making orthogonality assumptions are manifold. One reason is surely that they are considerably easier to optimize than more involved systems relying on non-orthogonal communication schemes (e.g. \cgls{snd}). The algorithm we develop here is well suited for exactly these more involved resource allocation problems.
For further references on resource allocation problems please refer to \cref{sec:RA}.

\subsection{Major Contributions}
The key contributions of this paper are the following:
\begin{itemize}
	\item The concept of essential feasibility is introduced and its relevance for resource allocation problems is shown with a simple example. The numerical problems of conventional global optimization algorithms are discussed and the \cgls{sit} scheme is introduced as a remedy for the these issues.
	\item We extend the \cgls{sit} algorithm as developed in \cite{Tuy2016} into an optimization framework able to deal with fractional objectives, non-convex variables, and multiple constraints. Specifically, we design an adaptive \cgls{bb} procedure that only branches over the global variables
		and leverages the power of commercially available state-of-the-art solvers to handle the non-global variables.
		To this end, we identify two different sets of sufficient conditions on the optimization problem. Fractional objectives are directly treated within the developed algorithm making the use of Dinkelbach's iterative algorithm obsolete.
	\item The framework is applied to optimize the throughput and energy efficiency of a \cgls{mwrc}.
		Extensive numerical results show a speed-up over state-of-the-art algorithms of four orders of magnitude for throughput maximization and almost three orders of magnitude over Dinkelbach's Algorithm for \cgls{ee} maximization.
	\item Code and data are made publicly available on GitHub \cite{github}. This allows other researchers to easily verify our results and adapt the \cgls{sit} algorithm for their own research.
\end{itemize}

\subsection{Notation \& Preliminaries} \label{sec:notation}
A vector $\vec x \in\mathds R^n$ with components $(x_1, \dots, x_n)$ is said to \emph{dominate} another vector $\vec y\in\mathds R^n$, i.e., $\vec y \le \vec x$, if $y_i \le x_i$ for all $i = 1, \dots, n$.
For $\vec a \le \vec b$, the set $[\vec a, \vec b] = \{ \vec x \,|\, \vec a \le \vec x \le \vec b\}$ is called a \emph{box}.
A function $f : \mathds R^n_{\ge0} \mapsto \mathds R$ is \emph{increasing} if $f(\vec x') \le f(\vec x)$ whenever $\vec x' \le \vec x$, and \emph{decreasing} if $-f$ is increasing.
It is called \emph{mixed monotonic} if it is increasing in the variables $(x_i)_{i\in\mathcal I}$ and decreasing in $(x_i)_{i\in\{1, 2, \ldots, n\} \setminus \mathcal I}$ for some index set $\mathcal I$.
The functions $f_1(\vec x), \dots, f_n(\vec x)$ are called \emph{jointly mixed monotonic} if all functions are mixed monotonic with respect to the same index set $\mathcal I$.
A \emph{common minimizer (maximizer)} of the functions $f_1(\vec x), \dots, f_n(\vec x)$ over the set $\mathcal X$ is any $\vec x^\ast$ that satisfies $\vec x^\ast \in \bigcap_{i=1}^n \argmin_{\vec x \in \mathcal X} f_i(\vec x)$ ($\vec x^\ast \in \bigcap_{i=1}^n \argmax_{\vec x \in \mathcal X} f_i(\vec x)$). 

A set $\mathcal G\subseteq\mathds R^n_{\ge0}$ is said to be \emph{normal} if for $0 \le \vec x' \le \vec x$, $\vec x\in\mathcal G$ $\Rightarrow$ $\vec x'\in\mathcal G$, and \emph{normal in a box $[\vec a, \vec b]$} if the previous implication only holds for $\vec a \le \vec x' \le \vec x \le \vec b$.
A set $\mathcal H\subseteq\mathds R^n_{\ge0}$ is called \emph{conormal} if $\vec x + \mathds R_{\ge0}^n\subseteq\mathcal H$ whenever $\vec x\in\mathcal H$, and \emph{conormal in a box $[\vec a, \vec b]$} if for $\vec b \ge \vec x' \ge \vec x \ge \vec a$, $\vec x\in\mathcal H \Rightarrow \vec x'\in\mathcal H$ \cite[Sec.~11.1.1]{Tuy2016}.
Let $\mathcal A\subseteq\mathds R^n$ and $(\vec x, \vec y)\in\mathcal A$. Then, $\proj_{\vec x} \mathcal A = \{ \vec x \,|\, (\vec x, \vec y)\in\mathcal C\ \text{for some}\ \vec y\}$, i.e., the projection of $\mathcal C$ onto the $\vec x$ coordinates;
$\diam\mathcal A$ is the \emph{diameter} of $\mathcal A$, i.e., the maximum distance between two points in $\mathcal A$;
and  $\mathcal A_{\tilde{\vec x}} = \{ \vec y | (\tilde{\vec x}, \vec y) \in \mathcal D\}$ is called the \emph{$\tilde{\vec x}$-section} of $\mathcal A$.

Finally, consider \cref{opt} and let its optimal value be $v\eqref{opt}$. To make the previously introduced notion of global and non-global variables more precise, split $\vec x$ into two vectors $\vec y = (x_i)_{i\in\mathcal I}$ and $\vec z = (x_i)_{i\in\{1, 2, \ldots, n\} \setminus \mathcal I}$ and consider a modified version of \cref{opt} where we only optimize over $\vec y$ for some fixed $\vec z$, i.e., $\max_{\vec y\in\mathcal C_{\vec z}} f(\vec y, \vec z)$. If there exists an algorithm to solve this optimization problem with computational complexity significantly less than required for solving the global part of \cref{opt},\footnote{Ideally, the computational complexity for solving the non-global part of \cref{opt} grows polynomially in the number of non-global variables. This is, for example, the case if \cref{opt} is linear in $\vec y$ \cite{Khachiyan1979,Karmarkar1984}. It also holds in many cases were \cref{opt} is convex in $\vec y$ \cite{Nesterov1994}. However, there exist convex optimization problems that are not solvable in polynomial time \cite[Sect.~5.4]{Ben-Tal2001} but still have computational complexity significantly less than general non-convex optimization problems \cite{Nemirovskii1983}. Nevertheless, it might be helpful to think of the terms ``global'' and ``non-global'' variables as synonyms for ``non-convex'' and ``convex'' variables, respectively.} the variables $\vec y$ and $\vec z$ are denoted as \emph{non-global} and \emph{global} variables, respectively.

\subsection{Outline}
The remaining part of this paper is organized as follows. In the next section, we formally state the considered optimization problem and motivate it in the context of resource allocation problems in wireless interference networks. \Cref{sec:math} introduces important mathematical preliminaries including the \cgls{sit} scheme that forms the basis of the proposed algorithm. These developed principles are then applied in \cref{sec:algo} to construct \cref{alg:sitbb}, one of the main contributions of this paper. In \cref{sec:numeval}, we apply the developed framework to a specific resource allocation problem that is used to benchmark our method against the state-of-the-art. Finally, we give our conclusions in \cref{sec:conclusions}.

\section{Problem statement}
We consider the following global optimization problem
\begin{optprob}
	& \underset{(\vec x, \vec\xi) \in \mathcal C}{\text{max}} 
	&& \frac{f^+(\vec x, \vec\xi)}{f^-(\vec x, \vec\xi)} \\
	& \text{s.\,t.}
	&& g_i^+(\vec x, \vec\xi) - g_i^-(\vec x) \le 0,\quad i = 1, 2, \dots, m,
	\label{opt:genRA}
\end{optprob}
with global variables $\vec x$ and non-global variables $\vec\xi$.
The functions $\{ g_i^-(\vec x) \}$ are required to have a common maximizer over every box $[\ubar{\vec x}, \bar{\vec x}] \subseteq\mathcal M_0$ with $\mathcal M_0$ being a box enclosing the $\vec x$ dimensions of $\mathcal C$, i.e., $\mathcal M^0 \supset \proj_{\vec x} \mathcal C$.
A sufficient condition for the existence of this common maximizer is that the functions $g_i^-(\vec x)$ are jointly mixed monotonic. This includes the case where all functions are either increasing or decreasing.
Further, we assume the functions $f^-(\vec x, \vec\xi)$, $g_i^+(\vec x, \vec\xi)$, $i = 1, \dots, m$, to be \cgls{lsc}, 
the functions $f^+(\vec x, \vec\xi)$, $g_i^-(\vec x)$, $i = 1, \dots, m$, to be \cgls{usc},
and, \cgls{wlog}, $f^-(\vec x, \vec\xi) > 0$.

The goal of this paper is to design a numerically stable \cgls{bb} procedure to solve \cref{opt:genRA} that preserves the computational complexity in the non-global variables $\vec\xi$. This requires additional assumptions to those stated above. Specifically, we have identified two different sets of technical requirements that are stated below. Both of these cases contain conditions that depend on a constant $\gamma$ which will hold the current best known value in the developed algorithm. We will discuss the domain of $\gamma$ after the definition of both cases below. 
 
\begin{assumption}[\cgls{dc} problems] \label{constrA}
	If $\mathcal C$ is a closed convex set and $\gamma f^-(\vec x, \vec\xi) - f^+(\vec x, \vec\xi)$, $g_1^+(\vec x, \vec\xi)$, \ldots, $g_m^+(\vec x, \vec\xi)$ are jointly convex in $(\vec x, \vec\xi)$ for all $\gamma$, problem~\cref{opt:genRA} resembles a \cgls{dc} optimization problem but with fractional objective and additional non-\cgls{dc} variables.
\end{assumption}

\begin{assumption}[Separable problems] \label{constrB}
	Let $\mathcal C = \mathcal X \times \Xi$ such that $\vec x\in\mathcal X$ and $\vec\xi\in\Xi$ with $\Xi$ being a closed convex set, and let each function of $(\vec x, \vec\xi)$ be separable in the sense that $h(\vec x, \vec\xi) = h_x(\vec x) + h_\xi(\vec\xi)$.
	Further, let the functions $\gamma f^-_\xi(\vec\xi) - f^+_\xi(\vec\xi)$, $g_{1,\xi}^+(\vec\xi)$, \dots, $g_{m,\xi}^+(\vec\xi)$ be convex in $\vec\xi$ for all $\gamma$, and let the functions $\gamma f^-_x(\vec x) - f^+_x(\vec x)$, $g_{1,x}^+(\vec x)$, \dots, $g_{m,x}^+(\vec x)$ have a common minimizer over $\mathcal X \cap \mathcal M$ for every box $\mathcal M\subseteq\mathcal M_0$ and all $\gamma$.
	Finally, let the function $\gamma f^-_x(\vec x) - f^+_x(\vec x)$ be either increasing for all $\gamma$ with $\mathcal X$ being a closed normal set in some box, or decreasing for all $\gamma$ with $\mathcal X$ being a closed conormal set in some box.
\end{assumption}

\begin{remark}\label{rem:preferCaseB}
	Separable problems often lead to linear auxiliary optimization problems with less variables instead of, typically, convex problems for \cref{constrA}. They usually have lower computational complexity than \cgls{dc} problems and, thus, if the problem at hand falls into both cases, it is usually favorable to consider it as a separable problem.
\end{remark}

With both cases defined we can continue our discussion of $\gamma$. 
First, observe that $\gamma$ only appears as a factor to $f^-(\vec x, \vec\xi)$.
Thus, its value is only relevant if \cref{opt:genRA} is a fractional program, i.e., if $f^-(\vec x, \vec\xi)$ is not constant.
In that case, the only relevant property of $\gamma$ is its sign and whether it may change during the algorithm. For example, in \cref{constrA} the function $\gamma f^-_\xi(\vec\xi) - f^+_\xi(\vec\xi)$ is convex if $f^+_\xi(\vec\xi)$ is concave and $\gamma f^-_\xi(\vec\xi)$ is convex. The latter is the case if $\gamma\ge0$ and $f^-_\xi(\vec\xi)$ is convex, or if $\gamma\le0$ and $f^-_\xi(\vec\xi)$ is concave. Thus, in most cases, we should ensure that the sign of $\gamma$ is constant. 
In general, $\gamma$ may take values between some $\gamma_0$ and $v(\ref{opt:genRA}) + \eta$ for some small $\eta > 0$. The lower end of the range $\gamma_0$ is either
the objective value of \cref{opt:genRA} for some preliminary known nonisolated feasible point $(\vec x, \vec\xi)$ or an arbitrary value satisfying $\gamma_0 \le \frac{f^+(\vec x, \vec\xi)}{f^-(\vec x, \vec\xi)}$ for all feasible $(\vec x, \vec\xi)$. This implies, e.g., that 
$\gamma$ is non-negative if $f^+(\vec x, \vec\xi)$ is non-negative.
Otherwise, it might be necessary to find a nonisolated feasible point such that $f^+(\vec x, \vec\xi)\ge0$ or transform the problem.

\subsection{Application Example: Resource Allocation in Interference Channels} \label{sec:RA}
Determining an achievable rate region of a communication network usually involves two steps: first, characterizing the achievable rate region with information theoretical tools, and, second, finding Pareto-optimal resource allocations. The corresponding optimization problem in many Gaussian interference networks is
\begin{optprob} \label{opt:exRA}
	& \underset{\vec p, \vec R}{\text{max}} 
	&& f(\vec p, \vec R) \\
	& \text{s.\,t.}
	&& \vec a_i^T \vec R \le \log\!\left( 1 + \frac{\vec b_i^T \vec p}{\vec c_i^T \vec p + \sigma_i} \right),\quad i = 1, \ldots, n \\
	&&& \vec R \ge 0,\quad \vec p\in[\vec 0, \vec P]
\end{optprob}
for some performance function $f(\vec p, \vec R)$ and positive vectors $\vec a_i, \vec b_i, \vec c_i \ge 0$, $i = 1, \ldots, n$.
The optimization variables $\vec p$ are the allocated transmit powers and $\vec R$ are the achievable transmission rates for asymptotically error-free communication. Usually, the vectors $\vec b_i$ and $\vec c_i$ represent the effective channel gain, $\sigma_i$ is the variance of the Gaussian noise observed at receiver $i$, $\vec a_i$ is a sparse vector where the non-zeros entries are small integers (mostly ones), and $\vec P$ are the maximum transmit powers.  Applications of this model include multi-cell communication systems \cite{Bjornson2013}, heterogeneous dense small-cell networks \cite{wcnc18}, cognitive radio \cite{Haykin2005}, and \cgls{dsl} systems \cite{Yu2002}.
For the sake of simplicity, we assume the \gls{sinr} in the \cgls{rhs} expressions of the constraints to be linear fractions of the transmit powers. We note that this is not always the case and that our framework is not limited to this case.

The feasible set of \cref{opt:exRA} belongs to the class of considered problems. We identify the global variables as $\vec p$ and the non-globals as $\vec R$. The rate constraints are equivalent to
\begin{equation*}
	\vec a_i^T \vec R + \log\!\left( \vec c_i^T \vec p + \sigma_i \right) - \log\!\left( \left( \vec b_i^T + \vec c_i^T \right) \vec p + \sigma_i \right) \le 0.
\end{equation*}
Since $\vec a_i, \vec b_i, \vec c_i\ge0$ for all $i$, the linear function $\vec a_i \vec R$ and the $\log$-functions are increasing. Thus, we can identify $g^+_i(\vec p, \vec R)$ and $g^-_i(\vec p)$ as:
\begin{equation}
\begin{aligned}
	g^+_i(\vec p, \vec R) &\coloneqq \vec a_i^T \vec R - \log\!\left(\left( \vec b_i^T + \vec c_i^T \right) \vec p + \sigma_i \right) \\
	g^-_i(\vec p) &\coloneqq -\log\!\left(  \vec c_i^T \vec p + \sigma_i \right).
\end{aligned}
\label{eq:identification1}
\end{equation}
The functions $g_i^-(\vec p)$ are decreasing, and, thus, are jointly maximized over the box $[\ubar{\vec p}, \bar{\vec p}]$ by $\ubar{\vec p}$. Further, $g^+_i(\vec p, \vec R)$ is separable in $\vec p$ and $\vec R$ with $g^+_{i, R}(\vec R)$ being linear and $g^+_{i, p}(\vec p)$ decreasing and convex in $\vec p$. Thus, depending on $f(\vec p, \vec R)$ problem \cref{opt:exRA} qualifies for both, \cref{constrA,constrB}.

The complete Pareto boundary is characterized by all solutions to the \cgls{mop} \cref{opt:exRA} with the vector objective $f(\vec p, \vec R) = [ R_1, R_2, \ldots, R_K ]$ \cite{Zadeh1963}. Several approaches exist to transform this into a scalar optimization problem. Two very popular are the scalarization and the rate profile approach\footnote{This approach is also known as rate balancing \cite{Jorswieck2002}.} \cite{Zhang2010a}.

\paragraph{Weighted Sum Rate}\label{sec:wsrm} In the scalarization approach, the weighted sum of the objectives is optimized, i.e., $f(\vec p, \vec R) = f^+(\vec p, \vec R) \coloneqq \vec w^T \vec R$. Obviously, with this objective, \cref{opt:exRA} belongs to the class of separable problems since \cref{constrB} is satisfied. Varying the weights between 0 and 1 with $\sum_k w_k = 1$ characterizes the convex hull of the Pareto boundary of the achievable rate region. This is by far the most widely used performance metric for power control in wireless communication systems due its clear operational meaning: with all weights $w_k = 1$, $f^+(\vec p, \vec R)$ is the total \emph{throughput} in the network. Moreover, in several networks this condition leads to the characterization of the stability region. In this queueing theoretic setting, choosing the weights proportional to the queue lengths prioritizes longer queues and stabilizes the network \cite{Tassiulas1992,Neely2003}.

Many instances of the weighted sum rate maximization problem considered in the literature are either convex optimization problems \cite{Tse1998,Jindal2005,Jorswieck2007a}, or have a rather good-natured rate region allowing to eliminate the rates $\vec R$ from the problem such that the resulting global optimization is only over the powers \cite{Qian2009,Jorswieck2010a,Bjornson2013}. If neither is the case, variable reduction techniques are necessary to keep the computational complexity at a reasonable level. In \cite{camsap17}, the problem at hand is transformed into a monotonic optimization problem \cite{Tuy2000} where the objective is a linear program with the transmit power as parameter. This approach is considerably slower, theoretically more involved, and less versatile than the proposed method.

\paragraph{Rate Profile Approach} 
The utility profile approach finds the intersection of a ray in the direction $\vec w$ and the Pareto boundary of the achievable rate region, i.e.,
\begin{optprob} \label{opt:exMaxmin}
	& \underset{\vec p, \vec R, t}{\text{max}} 
	&& t \\
	& \text{s.\,t.}
	&& \vec a_i^T \vec R \le \log\!\left( 1 + \frac{\vec b_i^T \vec p}{\vec c_i^T \vec p + \sigma_i} \right),\quad i = 1, \ldots, n \\
	&&& \vec R \ge t \vec w,\quad \vec p\in[\vec 0, \vec P]
\end{optprob}
for some $\vec w \ge 0$. \cGls{wlog} one can assume $\norm{\vec w} = 1$. By varying the direction of the ray, the complete Pareto boundary can be characterized. It is easily verified that \cref{opt:exMaxmin} satisfies \cref{constrB}.

Problem~\cref{opt:exMaxmin} is the epigraph form of \cref{opt:exRA} with $f(\vec p, \vec R) = \min_k \frac{R_k}{w_k}$ as long as $\vec w \neq \vec 0$. Since $f(\vec p, \vec R)$ is continuous and concave, it also satisfies \cref{constrB}. Besides reducing the number of variables, this reformulation also makes it apparent that the rate profile approach is equivalent to the \emph{weighted max-min fairness} performance function. For $\vec w = \vec 1$, it puts very strong emphasis on the weakest user and often results in low spectral efficiency. However, in some cases this approach might be computationally less challenging than the scalarization approach as we will discuss below.

\paragraph{Global Energy Efficiency} The \cgls{gee} is the most widely used metric to measure the network energy efficiency, a key performance metric in 5G and beyond networks \cite{Isheden2012,Zappone2016}.
It is defined as the benefit-cost ratio of the total network throughput and the associated power consumption, i.e.,
\begin{equation}
	\mathrm{GEE} = \frac{\sum_k R_k}{\vec\phi^T \vec p + P_c},
	\label{eq:gee}
\end{equation}
where $\vec\phi\ge 1$ are the inverses of the power amplifier efficiencies and $P_c$ is the total circuit power necessary to operate the network. Using the \cgls{gee} as performance metric results in a fractional programming problem \cite{Charnes1962,dinkelbach1967,Schaible1983,Schaible1993}. Usually, it is solved iteratively with Dinkelbach's Algorithm \cite{dinkelbach1967} where, in each iteration, an auxiliary optimization problem is solved. This auxiliary problem generally has very similar properties to the weighted sum rate maximization problem, i.e., for interference networks it is a global optimization problem with exponential complexity. This inner problem must be solved several times with high numerical accuracy for the convergence guarantees of Dinkelbach's Algorithm to hold \cite{Zappone2017}.

Instead, our proposed algorithm allows to solve these fractional programs directly resulting in significantly lower complexity. Consider the generic resource allocation problem \cref{opt:exRA} with objective \cref{eq:gee}. We identify $f^+(\vec p, \vec R) = \sum_k R_k$ and $f^-(\vec p, \vec R) = \vec\phi^T \vec p + P_c$ and observe that both are linear functions. Hence, $\gamma f^-(\vec x, \vec\xi) - f^+(\vec x, \vec\xi)$ is convex for all $\gamma$ and \cref{constrA} is satisfied. Of course, $f^+(\vec p, \vec R)$ and $f^-(\vec p, \vec R)$ are also separable, and because the \cgls{gee} is non-negative, $\gamma\ge0$ and $\gamma f_p^-(\vec p)$ is increasing. But since $g_{i,p}^+(\vec p)$ is decreasing, the functions $\gamma f_p^-(\vec p)$, $g_{1,p}^+(\vec p)$, \dots, $g_{m,p}^+(\vec p)$ do not have a common minimizer over $\mathcal X\cap\mathcal M$ for any box $\mathcal M$ and \cref{constrB} does not apply. However, observe that instead of the choice in \cref{eq:identification1}, we can also identify $g^+_i(\vec p, \vec R)$ and $g^-_i(\vec p)$ as
\begin{equation}
\begin{aligned}
	g^+_i(\vec p, \vec R) &\coloneqq \vec a_i^T \vec R + \log\!\left( \vec c_i^T \vec p + \sigma_i \right) \\
	g^-_i(\vec p) &\coloneqq \log\!\left( \left( \vec b_i^T + \vec c_i^T \right) \vec p + \sigma_i \right).
\end{aligned}
\label{eq:identification2}
\end{equation}
The functions $g^+_i(\vec p, \vec R)$ are separable in $\vec p$ and $\vec R$, and $g^+_{i,p}(\vec p)$ and $g^-_i(\vec p)$ are increasing in $\vec p$. Hence, $\gamma f_p^-(\vec p)$, $g_{1,p}^+(\vec p)$, \dots, $g_{m,p}^+(\vec p)$ have a common minimizer over $\mathcal X\cap\mathcal M$ for every box $\mathcal M$ and \cref{constrB} applies.
In general, \cref{eq:identification1} will result in tighter bounds and, thus, faster convergence. On the other hand, \cref{constrB} has significantly lower numerical complexity than \cref{constrA} (cf.~\cref{rem:preferCaseB}) which should compensate for this drawback. However, a final assessment is only possible through numerical experimentation which will be carried out in \cref{sec:numeval}.

\paragraph{Energy Efficiency Region}
Similar to the \cgls{gee}, the individual energy efficiency of link $k$ is defined as the benefit-cost ratio of the link's throughput divided by the link's power consumption, i.e., $\mathrm{EE}_k = \frac{R_k}{\phi_k p_k + P_{c,k}}$. Analogue to the achievable rate region, there is also an achievable \cgls{ee} region whose Pareto boundary is the solution to the \cgls{mop} \cref{opt:exRA} with objective $f(\vec p, \vec R) = [ \mathrm{EE}_1, \ldots, \mathrm{EE}_K]$ \cite{Zappone2015}. The same approaches as discussed earlier can be used to transform this \cgls{mop} into a scalar optimization problem. In this case, the rate profile (or rather utility profile) approach should be favored because the scalarization approach leads to a sum-of-ratios problem. The sum-of-ratios problem is one of the most difficult fractional programs and known to be essentially $\mathcal{NP}$-hard \cite{Freund2001,Schaible2003}. While for other \cgls{ee} problems (e.g., the \cgls{gee}) first-order optimal solutions are usually observed to be globally optimal \cite{Zappone2017} this does not hold for the sum-of-ratios case \cite{wcnc18}.

Instead, with the utility profile approach the objective in \cref{opt:exRA} is the weighted minimum \cgls{ee} $f(\vec p, \vec R) = \min_k \frac{\mathrm{EE}_k}{w_k}$. This is, as is the sum-of-ratios problem, a multi-ratio optimization problem but numerically less challenging. Another benefit of using the utility profile approach over the scalarization approach is that the \cgls{ee} region is non-convex and the scalarization approach only obtains the convex hull of the Pareto region \cite{Zappone2017}. The general approach to solving such an optimization problem is to use the Generalized Dinkelbach's Algorithm \cite{Crouzeix1985,Zappone2015,Zappone2017}. As in the \cgls{gee} case, our proposed method solves this problem with much lower complexity. However, some minor modifications of the algorithm are necessary to deal with this kind of problem. Please refer to \cref{sec:ext:min} for more details. 

\paragraph{Proportional Fairness} Another well known performance function is proportional fairness where the objective is the product of the rates, i.e., $\prod_k R_k$ \cite{Kelly1998}. The operating point achieved by this metric is usually almost as fair as the one obtained by max-min fairness but achieves significantly higher throughput. Observe that the objective is log-concave. Thus, we can determine the proportional fair operating point by solving \cref{opt:exRA} with objective $f(\vec p, \vec R) = \sum_k \log(R_k)$.

\section{Robust Approach to Global Optimization} \label{sec:math}
In this section, we introduce some mathematical preliminaries essential for developing the proposed algorithm. This is done by means of a simpler problem than \cref{opt:genRA} to ease the exposition. Specifically,
we consider the optimization problem
\begin{equation}
	\underset{\vec x\in [\vec a, \vec b]}{\text{max}} \enskip
	f(\vec x)
	\quad\text{s.\,t.}\quad
	g_i(\vec x) \le 0,\enskip i = 1, 2, \dots, m
\refstepcounter{optimizationproblem}\tag{P\theoptimizationproblem}
\label{opt:generalNC}
\end{equation}
where $f, g_1, g_2, \dots, g_m$ are non-convex continuous real-valued functions and $\vec a, \vec b$ are real-valued vectors satisfying $\vec a \le \vec b$. This is a general non-convex optimization problem with possibly quite complicated feasible set.

\subsection{The Issue with $\varepsilon$-Approximate Solutions}
Most current solution methods for this problem are devised to compute a solution $\bar{\vec x}(\varepsilon)$ of the $\varepsilon$-relaxed problem
\begin{equation}
	\underset{\vec x\in [\vec a, \vec b]}{\text{max}} \enskip
	f(\vec x)
	\quad\text{s.\,t.}\quad
	g_i(\vec x) \le \varepsilon,\enskip i = 1, 2, \dots, m,
\refstepcounter{optimizationproblem}\tag{P\theoptimizationproblem}
\end{equation}
This solution $\bar{\vec x}(\varepsilon)$ is usually accepted as an \emph{$\varepsilon$-approximate optimal solution} to \cref{opt:generalNC} since $\bar{\vec x}(\varepsilon)$ is almost feasible for small $\varepsilon > 0$ and tends to a feasible solution as $\varepsilon\rightarrow 0$. Since $f(\bar{\vec x}(\varepsilon))$ also tends to the optimal value $v(\ref{opt:generalNC})$, $f(\bar{\vec x}(\varepsilon))$ should be close to $v(\ref{opt:generalNC})$ for a sufficiently small $\varepsilon = \varepsilon_0$. The problem with this approach is that $\varepsilon_0$ is, in general, unknown and hard to determine. Thus, the obtained solution and $f(\bar{\vec x}(\varepsilon)$) can be quite far away from the true optimal value even for a small $\varepsilon$ \cite{Tuy2005a}. This is apparent from the following example.

\begin{figure}
	\centering
	\tikzsetnextfilename{leakage}
	\begin{tikzpicture}
		\def\leakage{8.99}
		\begin{axis} [
			xmin = .8,
			xmax = 5,
			ymin = .8,
			ymax = 5,
			xtick = {0.8,5},
			ytick = {5},
			xlabel = {$p_1$},
			ylabel = {$p_2$},
			ylabel near ticks,
			ylabel style = {rotate=-90},
			major tick length = 0,
			smooth,
			no markers,
			samples=50,
			domain = 1:5,
			thick,
			nodes near coords align={anchor = south west},
			nodes near coords style={inner xsep = 0},
			width=\axisdefaultwidth,
			height=.8*\axisdefaultheight,
		]
			\addplot[] expression { (-x-2+2*\leakage)/(x+2) };
			\addplot[] expression { -x + 6 };

			\addplot[draw=none,name path=L] expression { (-x-2+2*\leakage)/(x+2) } \closedcycle;
			\addplot[draw=none,name path=Q] expression { -x + 6 } \closedcycle;
			\addplot[blue!40] fill between[of=Q and L,split, every even segment/.style={fill=none}];

			\addplot[nodes near coords, only marks, point meta=explicit symbolic, mark options={mark=*, scale=1.0, fill=black, draw=black, mark size=1pt}] coordinates {
				(4.00665, 1.99335) [$\vec{p}^\ast$ / $\bar{\vec p}(\varepsilon_2)$]
				(0.995843, 5) [$\bar{\vec p}(\varepsilon_1)$]
			};

			\coordinate (p1) at (axis cs:2.75,3.25);
			\node [xshift = 15pt, yshift = 10pt, inner sep = 2pt, anchor=west] (qos) at (p1) {QoS};
			\draw [-{Latex[open,bend]}] (qos.west) to[bend right] (p1);

			\coordinate (p2) at (axis cs:3,2.596);
			\node [xshift = -15pt, yshift = -10pt, inner sep = 2pt, anchor=east] (leakage) at (p2) {Leakage};
			\draw [-{Latex[open,bend]}] (leakage.east) to[bend right] (p2);

			\coordinate (p2) at (axis cs:4.5,1.5);
			\node [xshift = -5pt, yshift = -10pt, inner sep = 2pt, anchor=east] (leakage) at (p2) {Feasible Set};
			\draw [-{Latex[open,bend]}] (leakage.east) to[bend right] ($(p2)+(10pt,0pt)$);
		\end{axis}
	\end{tikzpicture}
	\vspace{-2ex}
	\caption{Feasible set of \cref{ex:leakage} with optimal solution $\vec p^\ast$ and $\varepsilon$-approximate solutions $\bar{\vec p}(\varepsilon_1)$ and $\bar{\vec p}(\varepsilon_2)$ for $\varepsilon_1 = 10^{-3}$ and $\varepsilon_2 = 10^{-4}$. The $\varepsilon_1$-approximate solution is quite far away from the true optimum.}
	\label{fig:leakage}
\end{figure}
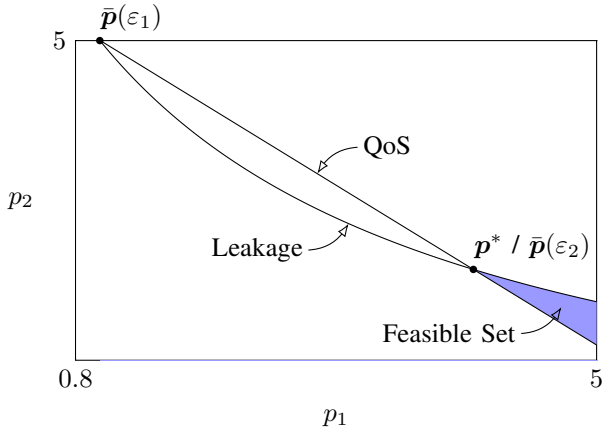

\begin{example}[Issues with $\varepsilon$-approximate solutions] \label{ex:leakage}
	Consider a Gaussian 2-user \cgls{siso} \cgls{mac} with channel gains $h_1$ and $h_2$, transmit powers $p_1\le P_1$ and $p_2\le P_2$, and a minimum total throughput of $Q$, i.e., $\log_2(1 + \abs{h_1}^2 p_1 + \abs{h_2}^2 p_2) \ge Q$. Transmitter $i$, $i = 1, 2$, is eavesdropped by a single antenna adversary over a channel $g_i$. The eavesdroppers are only able to overhear one of the transmitters and do not cooperate with each other \cite{Barros2006}. The total information leakage to the eavesdroppers is limited by $L$, i.e., $\log_2(1+\abs{g_1}^2 p_1) + \log_2(1 + \abs{g_2}^2 p_2) \le L$. The transmit power of transmitter~1 should be minimized without violating these constraints. The resulting feasible set for $\abs{h_1}^2 = \abs{h_2}^2 = 10$, $\abs{g_1}^2 = \frac{1}{2}$, $\abs{g_2}^2 = 1$, $Q = \log_2(61)$, and $L = \log_2(8.99)$ is shown in \cref{fig:leakage}. The true optimum solution is $\vec p^\ast = (4.00665, 1.99335)$ with $f(\vec{p}^\ast) = 4.00665$,	while the $\varepsilon$-approximate solution for $\varepsilon_1 = 10^{-3}$ is $\bar{\vec p}(\varepsilon_1) = (0.995843, 5)$ with $f(\bar{\vec p}(\varepsilon_1)) = 0.995843$. This is, obviously, quite far away from both, the optimal solution and value. Instead, the $\varepsilon$-approximate solution obtained for $\varepsilon_2 = 10^{-4}$ is $\bar{\vec p}(\varepsilon_2) = (4.00541, 1.99417)$ with $f(\bar{\vec p}(\varepsilon_2)) = 4.00541$.
%
%
%
%
%
%
%
\end{example}

Thus, even for well behaved problems the $\varepsilon$-relaxation approach might fail. A slightly better approach to the approximate optimal solution of \cref{opt:generalNC} is that of an \emph{$\eta$-optimal solution}. A feasible vector $\bar{\vec x}$ is called $\eta$-optimal solution if it satisfies $f(\bar{\vec x}) \ge f(\vec x) - \eta$ for all feasible $\vec x$. The downside of this approach is that it does not converge in finitely many steps if the optimal solution is an isolated feasible point, i.e., a point at the center of a ball containing no other feasible points \cite{Tuy2005a}.

\begin{example}[Isolated optimal solution]
	Consider \cref{ex:leakage} again. With $L = \log_2(9)$ there exists an isolated feasible point $\vec p = (1, 5)$ (close to $\bar{\vec p}(\varepsilon_1)$ in \cref{fig:leakage}) that also happens to be the optimal solution. However, apart from the algorithmic difficulties in computing it, it might also be unstable under small pertubations of the data and is, thus, quite impractical from an engineering point of view.
\end{example}

\subsection{$\varepsilon$-Essential Feasibility}
A common approach to this dilemma is to assume that the feasible set $\mathcal F$ is robust, i.e., it satisfies $\mathcal F^\ast = \closure(\interior\mathcal F)$ where $\closure$ and $\interior$ denote the closure and interior, respectively. Unfortunately, this condition is generally very hard to check, so that, in practice, we have to deal with feasible sets where we do not know a priori whether they are robust or not.

This motivates the concept of $\varepsilon$-essential optimality developed by Tuy \cite{Tuy2005a,Tuy2009,Tuy2016}.  
A solution $\vec x^\ast\in\mathcal F^\ast$ is called \emph{essential optimal solution} of \cref{opt:generalNC} if
$f(\vec x^\ast) \ge f(\vec x)$ for all $x\in\mathcal F^\ast$.
A point $\vec x\in[\vec a, \vec b]$ satisfying $g_i(\vec x) \le -\varepsilon$ for all $i$ and some $\varepsilon > 0$ is called \emph{$\varepsilon$-essential feasible} and a solution of \cref{opt:generalNC} is said to be \emph{essential $(\varepsilon, \eta)$-optimal} if it satisfies
\begin{equation}
	f(\vec x^\ast) + \eta \ge \sup\{ f(\vec x) | \vec x\in[\vec a, \vec b],\ \forall i : g_i(\vec x) \le -\varepsilon \},
\end{equation}
for some $\eta > 0$.
Clearly, for $\varepsilon, \eta \rightarrow 0$ an essential $(\varepsilon, \eta)$-optimal solution is a nonisolated feasible point which is optimal.

\subsection{Successive Incumbent Transcending Scheme}
The robust approach to global optimization employed here uses the \cgls{sit} scheme in \cref{alg:sit} to generate a sequence of nonisolated feasible solutions converging to an essential optimal solution of \cref{opt:generalNC}.
The core problem in the \cgls{sit} scheme is, given a real number $\gamma$, to check whether \cref{opt:generalNC} has a nonisolated feasible solution $\vec x$ satisfying $f(\vec x) \ge \gamma$, or, else, establish that no such $\varepsilon$-essential feasible $\vec x$ exists.
Given that this subproblem is solved within finitely many steps, \cref{alg:sit} converges to the global optimal solution within finitely many iterations.
Apart from the improved numerical stability and convergence, the \cgls{sit} algorithm has another very desirable feature: it provides a good nonisolated feasible (but possibly suboptimal) point even if terminated prematurely. 
Instead, conventional algorithms usually outer approximate the solution rendering intermediate solutions almost useless.
\begin{algorithm}
	\def\tuyref{\cite[Sect.~7.5.1]{Tuy2016}}
	\caption{\cGls{sit} Algorithm \tuyref.}\label{alg:sit}
	\small
	\centering
	\begin{minipage}{\linewidth-1em}
		\begin{enumerate}[label=\textbf{Step \arabic*},ref=Step~\arabic*,start=0,leftmargin=*]
			\item \label{alg:sit:step0} Initialize $\bar{\vec x}$ with the best known nonisolated feasible solution and set $\gamma = f(\bar{\vec x}) + \eta$; otherwise do not set $\bar{\vec x}$ and choose  $\gamma \le f(\vec x)$ $\forall \vec x \in \mathcal F$.
			\item\label{step1} Check if \cref{opt:generalNC} has a nonisolated feasible solution $\vec x$ satisfying $f(\vec x) \ge \gamma$; otherwise, establish that no such $\varepsilon$-essential feasible $\vec x$ exists and go to \ref{step3}.
			\item Update $\bar{\vec x} \gets \vec x$ and $\gamma \gets f(\bar{\vec x}) + \eta$. Go to \ref{step1}.
			\item\label{step3} Terminate: If  $\bar{\vec x}$ is set, it is an essential $(\varepsilon, \eta)$-optimal solution; else Problem \cref{opt:generalNC} is $\varepsilon$-essential infeasible.
		\end{enumerate}
	\end{minipage}
\end{algorithm}

Consider the optimization problem
\begin{equation}
	\underset{\vec x\in [\vec a, \vec b]}{\text{min}} \enskip
	\underset{i = 1, 2, \dots, m}{\text{max}}\
	g_i(\vec x)
	\quad\text{s.\,t.}\quad
	f(\vec x) \ge \gamma
\refstepcounter{optimizationproblem}\tag{P\theoptimizationproblem}
\label{opt:dualNC}
\end{equation}
where we interchanged objective and constraints of \cref{opt:generalNC}.
Very often $f(x)$ has, or could be modified easily to have, nice properties like being concave or increasing, such that the feasible set of \cref{opt:dualNC} is \emph{nice}, i.e., it is robust and a 
feasible point can be computed efficiently using an adaptive \cgls{bb} procedure \cite[Prop.~6.2]{Tuy2016}.
The following proposition, which is an adapted version of \cite[Prop.\,7.13]{Tuy2016}, establishes a duality between \cref{opt:generalNC} and \cref{opt:dualNC} in the sense that
the feasibility problem in \ref{step1} of \cref{alg:sit} is equivalent to solving \cref{opt:dualNC}.
\begin{proposition} \label{prop:duality}
	For every $\varepsilon>0$, the $\varepsilon$-essential optimal value of \cref{opt:generalNC} is less than $\gamma$ if the optimal value of \cref{opt:dualNC} is greater than $-\varepsilon$.
\end{proposition}
\def\tuyref{\cite[Prop.~7.13]{Tuy2016}}
\begin{IEEEproof}[Proof (adapted from \tuyref)]
	If the optimal value of \cref{opt:dualNC} is greater than $-\varepsilon$, then any $\vec x \in[\vec a, \vec b]$ such that $g_i(\vec x) \le -\varepsilon$, for all $i = 1, 2, \dots, m$, must satisfy $f(\vec x) < \gamma$. Hence, by the compactness of the feasible set of \cref{opt:generalNC}, $\max\{ f(\vec x) | \vec x\in[\vec a, \vec b],\  \forall i : g_i(\vec x) \le -\varepsilon \} < \gamma$.
\end{IEEEproof}

Observe that every point $\vec x'$ in the feasible set of \cref{opt:dualNC} with objective value less or equal than zero is also a feasible point of \cref{opt:generalNC} with objective value better than $\gamma$. Thus, we can solve \cref{opt:generalNC} sequentially by solving \cref{opt:dualNC} with a \cgls{bb} method. Each time this \cgls{bb} algorithm finds a feasible point $\vec x'$ with objective value less or equal than zero (in \cref{opt:dualNC}), the current best value $\gamma$ is updated with the objective value of $\vec x'$ in the original problem \cref{opt:generalNC} (plus the tolerance $\eta$). Then, the \cgls{bb} solver continues solving \cref{opt:dualNC} with updated feasible set until it either finds a new point to update $\gamma$ or establishes that no solution to \cref{opt:dualNC} with objective value less or equal than $-\varepsilon$ exists. By virtue of \cref{prop:duality}, the last feasible point $\vec x'$ that was used to update $\gamma$ is an $(\varepsilon, \eta)$-optimal solution of \cref{opt:generalNC}. This observation will be formalized in \cref{prop:bb}.

In the next section, we return our attention to \cref{opt:genRA} and use \cref{alg:sit,prop:duality} to solve it globally.

\section{Robust Global Resource Allocation} \label{sec:algo}
We now apply the theory developed in the previous section to the solution of \cref{opt:genRA}. Recall that a core idea is to exchange objective and constraints to obtain an optimization problem that is considerably easier to solve with a \cgls{bb} procedure than the original problem. This is mainly due to nice structural properties of the dual feasible set which facilitates an easy implementation of the feasibility checks required in \cgls{bb} methods and has no isolated points. These points are hard to compute and can lead to numerical instabilities.

Interchanging objective and constraints in \cref{opt:genRA} leads to
\begin{optprob*}
	& \underset{(\vec x, \vec\xi) \in \mathcal C}{\text{min}} 
	&& \underset{i = 1, 2, \dots, m}{\text{max}} \left( g_i^+(\vec x, \vec\xi) - g_i^-(\vec x) \right) \\
	& \text{s.\,t.}
	&& \frac{f^+(\vec x, \vec\xi)}{f^-(\vec x, \vec\xi)} \ge \gamma
\end{optprob*}
or, equivalently,
\begin{optprob}
	& \underset{(\vec x, \vec\xi) \in \mathcal C}{\text{min}} 
	&& \underset{i = 1, 2, \dots, m}{\text{max}} \left( g_i^+(\vec x, \vec\xi) - g_i^-(\vec x) \right) \\
	& \text{s.\,t.}
	&& \gamma f^-(\vec x, \vec\xi) - f^+(\vec x, \vec\xi) \le 0
	\label{opt:dualRA}
\end{optprob}
since $f^-(\vec x, \vec\xi) > 0$ by assumption. The feasible set $\mathcal D$ of \cref{opt:dualRA} is nice as long as the conditions in \cref{constrA} or~\ref{constrB} is satisfied. This is formally established in the proposition below.

\begin{proposition} \label{prop:robust}
	The feasible set of \cref{opt:dualRA} does not contain any isolated points if the conditions in \cref{constrA} or~\ref{constrB} are satisfied.
\end{proposition}
\begin{IEEEproof}
	Please refer to the appendix.
\end{IEEEproof}

We now design a \cgls{bb} procedure to solve \cref{opt:dualRA}. Together with the SIT scheme in \cref{alg:sit} this will result in a method to solve \cref{opt:genRA} and is stated in \cref{alg:sitbb}. The core idea of \cgls{bb} is to relax the feasible set and subsequently partition it such that lower bounds on the objective value can be determined easily. In our case, a rectangular subdivision procedure is a reasonable choice due to the required existence of a common maximizer of $\{g_i^-(\vec x)\}$ over every box in the domain of \cref{opt:dualRA}.

Since the \cgls{bb} procedure is supposed to only operate on the global variables $\vec x$, it partitions the $\vec x$-dimensions of $\mathcal D$ successively into boxes $\{ \mathcal M_i \}$. Specifically, in iteration $k$ the algorithm selects a box $\mathcal M^k = [\vec p^k, \vec q^k]$ with the lowest bound and bisects it via $(\vec v^k, j_k)$, i.e., $\mathcal M^k$ is replaced by
\begin{equation}
	\begin{aligned}
		\mathcal M^k_- &= \{ \vec x \,|\, p^k_{j_k} \le x_{j_k} \le v^k_{j_k},\ p^k_i \le x_i \le q^k_i\ (i\neq j_k) \} \\
		\mathcal M^k_+ &= \{ \vec x \,|\, v^k_{j_k} \le x_{j_k} \le q^k_{j_k},\ p^k_i \le x_i \le q^k_i\ (i\neq j_k) \}.
	\end{aligned}
	\label{eq:bisect}
\end{equation}
For each box $\mathcal M_i$, a lower bound $\beta(\mathcal M_i)$ for \cref{opt:dualRA} with additional constraint $\vec x\in\mathcal M_i$ is computed. We will discuss the computation of $\beta(\mathcal M_i)$ in the next subsection. For now it suffices to assume that it generates two points $\vec x^k\in \mathcal M^k$, $\vec y^k\in \mathcal M^k$ satisfying
\begin{multline}\label{eq:bbconv}
	(\vec x^k, \vec\xi^k)\in\mathcal D,\quad \underset{\vec\xi\in\mathcal D_{\vec y^k}}{\text{min}} g(\vec y^k, \vec\xi) - \beta(\mathcal M^k) \rightarrow 0\\ \text{as}\ \norm{\vec x^k - \vec y^k} \rightarrow 0
\end{multline}
for some $\vec\xi^k$ and with
$$g(\vec x, \vec\xi) = \operatorname{\text{max}}_{i = 1, 2, \dots, m} \left( g_i^+(\vec x, \vec\xi) - g_i^-(\vec x) \right).$$
Thus, we can employ an adaptive bisection that exhibits much faster convergence than the common exhaustive subdivision, and choose the bisection parameters in \cref{eq:bisect} as $\vec v^k = \frac{1}{2} ( \vec x^k + \vec y^k)$ and $j_k = \argmax_{j} \abs{y_j^k - x_j^k}$. For a formal proof of the convergence of such an adaptive \cgls{bb} procedure, we refer the interested reader to \cite[Prop.~6.2]{Tuy2016}.

\subsection{Bounding}
We now discuss the computation of lower bounds for \cref{opt:dualRA} that satisfy \cref{eq:bbconv}. First, we establish a lower bound on the objective of \cref{opt:dualRA}.
\begin{proposition} \label{prop:underestimator}
	Let $\bar{\vec x}^\ast_{\mathcal M}$ be a common maximizer of $\{g_i^-(\vec x)\}_{i=1, \dots, m}$ over the box $\mathcal M$.
	Then, \cref{opt:dualRA}'s objective is lower bounded over $\mathcal M$ by
	\begin{equation}
		\max_{i=1, 2, \dots, m} \left\{ g_i^+(\vec x, \vec\xi) - g_i^-(\bar{\vec x}^\ast_{\mathcal M}) \right\}
		\label{eq:underestimator}
	\end{equation}
	This bound is tight at $\bar{\vec x}^\ast_{\mathcal M}$.
\end{proposition}

\begin{IEEEproof}
Please refer to the appendix.
\end{IEEEproof}

With \cref{prop:underestimator} we can determine the lower bound $\beta(\mathcal M_i)$ as the optimal value of
\begin{optprob}
	& \underset{\vec x, \vec\xi}{\text{min}} 
	&&  \max_{i=1, 2, \dots, m} \left\{ g_i^+(\vec x, \vec\xi) - g_i^-(\bar{\vec x}^\ast_{\mathcal M_i}) \right\} \\
	& \text{s.\,t.}
	&& \gamma f^-(\vec x, \vec\xi) - f^+(\vec x, \vec\xi) \le 0 \\
	&&& (\vec x, \vec\xi) \in \mathcal C,\enskip \vec x\in \mathcal M_i.
	\label{opt:bound}
\end{optprob}
This is a convex optimization problem if the conditions in \cref{constrA} or~\ref{constrB} are satisfied and can be solved in polynomial time under very mild assumptions using standard tools \cite{Nesterov1994}. For \cref{constrA}, the objective of \cref{opt:bound} is a convex function because $g_i^-(\bar{\vec x}^\ast_{\mathcal M_i})$ is constant and the feasible set is convex (cf.\ \cref{prop:robust}). Thus, \cref{opt:bound} is convex given \cref{constrA}.
For \cref{constrB}, \cref{opt:bound} can be written as
\begin{optprob}
	& \underset{\vec x, \vec\xi}{\text{min}} 
	&&  \max_{i=1, 2, \dots, m} \left\{ g_{i,\xi}^+(\vec\xi) + g_{i,x}^+(\vec x) - g_i^-(\bar{\vec x}^\ast_{\mathcal M_i}) \right\} \\
	& \text{s.\,t.}
	&&  \gamma f_\xi^-(\vec\xi) - f_\xi^+(\vec\xi) + \gamma f_x^-(\vec x) - f_x^+(\vec x) \le 0 \\
	&&& \vec\xi \in \Xi,\enskip \vec x\in \mathcal X \cap \mathcal M_i.
	\label{opt:bound2}
\end{optprob}
Let $\ubar{\vec x}^\ast_{\mathcal M_i}$ be the common minimizer of $\gamma f^-_x(\vec x) - f^+_x(\vec x)$, $g_{1,x}^+(\vec x)$, \dots, $g_{m,x}^+(\vec x)$ over $\mathcal X\cap\mathcal M_i$.
Then \cref{opt:bound2} is equivalent to
\begin{optprob}
	& \underset{\vec\xi}{\text{min}} 
	&&  \max_{i=1, 2, \dots, m} \left\{ g_{i,\xi}^+(\vec\xi) + g_{i,x}^+(\ubar{\vec x}^\ast_{\mathcal M_i}) - g_i^-(\bar{\vec x}^\ast_{\mathcal M_i}) \right\} \\
	& \text{s.\,t.}
	&&  \gamma f_\xi^-(\vec\xi) - f_\xi^+(\vec\xi) + \gamma f_x^-(\ubar{\vec x}^\ast_{\mathcal M_i}) - f_x^+(\ubar{\vec x}^\ast_{\mathcal M_i}) \le 0 \\
	&&& \vec\xi \in \Xi.
	\label{opt:bound3}
\end{optprob}
It is easy to see that $\ubar{\vec x}^\ast_{\mathcal M_i}$ is the optimal solution of \cref{opt:bound2} since it jointly minimizes the objective and the first constraint. Problem~\cref{opt:bound3} is convex due to the assumptions on $\Xi$ and the remaining functions of $\vec\xi$ made in \cref{constrB}.

Finally, for each $\mathcal M^k$,  we identify the variables from \cref{eq:bbconv} as $\vec y^k = \bar{\vec x}^\ast_{\mathcal M^k}$ and $(\vec x^k, \vec\xi^k)$  as the optimal solution of \cref{opt:bound} or $\vec x^k = \ubar{\vec x}^\ast_{\mathcal M_i}$ if \cref{opt:bound3} is solved.

\subsection{The SIT Algorithm}
The original \cgls{sit} algorithm from \cite[Sect.~7.5.2]{Tuy2016} does not distinguish between global and non-global variables. We extent it such that branching is only performed over the global variables and state-of-the-art commercially available solvers can be used for the non-global variables. This preserves the computational complexity in the non-global variables, and increases computational performance and numerical accuracy compared to self-crafted algorithms due to the high maturity of these industry-grade solvers.
We also extend it to fractional objectives which removes the necessity for Dinkelbach's algorithm.

The \cgls{bb} procedure from the previous section solves \cref{opt:dualRA}, but that is not exactly what is required by the \cgls{sit} scheme. Instead, \cref{alg:sit} requires the implementation of
\begin{center}
	\small
	\begin{minipage}{\linewidth-2em}
		\begin{enumerate}[label=\textbf{Step \arabic*},ref=Step~\arabic*,start=1,leftmargin=*]
			\item\label{step1} Check if \cref{opt:genRA} has a nonisolated feasible solution $\vec x$ satisfying $f(\vec x) \ge \gamma$; otherwise, establish that no such $\varepsilon$-essential feasible $\vec x$ exists and go to \ref{step3}.
		\end{enumerate}
	\end{minipage}
\end{center}
This is accomplished by a modified version of the adaptive \cgls{bb} algorithm from the previous section. Consider the following proposition which is adapted from \cite[Prop.~7.14]{Tuy2016} and leverages the simple observation in \cref{prop:duality}.
\begin{proposition} \label{prop:bb}
	Let $\varepsilon> 0$ be given. Either $g(\vec x^k, \vec\xi^\ast) < 0$ for some $k$ and $\vec\xi^\ast$ or $\beta(\mathcal M^k) > -\varepsilon$ for some $k$. In the former case, $(\vec x^k, \vec\xi^\ast)$ is a nonisolated feasible solution of \cref{opt:generalNC} satisfying $\frac{f^+(\vec x^k, \vec\xi^\ast)}{f^-(\vec x^k, \vec\xi^\ast)}\ge\gamma$. In the latter case, no $\varepsilon$-essential feasible solution $(\vec x, \vec\xi)$ of \cref{opt:generalNC} exists such that $\frac{f^+(\vec x, \vec\xi)}{f^-(\vec x, \vec\xi)}\ge\gamma$.
\end{proposition}
\begin{IEEEproof}
	Straightforward adaption of \cite[Prop.~7.14]{Tuy2016}.
\end{IEEEproof}

Thus, an adaptive \cgls{bb} algorithm for solving \cref{opt:dualNC} with deletion criterion $\beta(M) > -\varepsilon$ and stopping criterion $\operatorname{\text{min}}_{\vec\xi\in\mathcal D_{\vec x^k}} g(\vec x^k, \vec\xi) < 0$ implements \ref{step1} in \cref{alg:sit}: In the first case of \cref{prop:bb} the incumbent feasible solution can be improved, in the latter case, if $\gamma =\frac{f^+(\bar{\vec x}, \bar{\vec\xi})}{f^-(\bar{\vec x}, \bar{\vec\xi})} + \eta$ for a given $\eta>0$ and a nonisolated feasible solution $(\bar{\vec x}, \bar{\vec\xi})$, the incumbent $(\bar{\vec x}, \bar{\vec\xi})$ is an essential $(\varepsilon, \eta)$-optimal solution of \cref{opt:generalNC}.

With the observation in \cref{prop:bb}, we can formally state the complete procedure for solving \cref{opt:genRA} with global optimality in \cref{alg:sitbb}.
It is initialized in Step~\ref{alg:sitbb:0} where an initial box $\mathcal M^0 = [\vec p^0, \vec q^0]$ is required that contains the $\vec x$-dimensions of $\mathcal C$, i.e.,
\begin{equation}
	p^0_i = \underset{(\vec x, \vec\xi)\in\mathcal C}{\text{min}}\enskip x_i 
	\qquad
	q^0_i = \underset{(\vec x, \vec\xi)\in\mathcal C}{\text{max}}\enskip x_i.
	\label{eq:M0}
\end{equation}
The set $\mathscr P_k$ contains new boxes to be examined in Step~\ref{alg:sitbb:1}, $\gamma$ holds the current best value adjusted by the tolerance $\eta$, and $\mathscr R$ holds all boxes that are not yet eliminated. In Step~\ref{alg:sitbb:1} the bound is computed for each box in $\mathscr P_k$. If it is less than $-\varepsilon$, the box may contain a nonisolated feasible solution with objective value greater than $\gamma$ and is added to $\mathscr R$. Then, in Step~\ref{alg:sitbb:3}, the box with the smallest bound is taken out of $\mathscr R$. If the point $\vec x^k$ attaining the bound is feasible in the original problem \cref{opt:genRA}, it is a nonisolated feasible point and needs to be examined further: if the objective value for $\vec x^k$ is greater than the current best value $\gamma-\eta$, $\vec x^k$ is the new current best solution and $\gamma$ is updated accordingly in Step~\ref{alg:sitbb:4}.
Irrespective of $\vec x^k$'s feasibility, the box selected in Step~\ref{alg:sitbb:3} is bisected via $(j_k, \vec v^k)$ adaptively. These new boxes are then passed to Step~\ref{alg:sitbb:1} and the algorithm repeats until $\mathscr R$ holds no more boxes which is checked in Step~\ref{alg:sitbb:2}. Convergence of the algorithm is stated formally in the theorem below.

\begin{theorem} \label{thm:convergence}
	\Cref{alg:sitbb} converges in finitely many steps to the $(\varepsilon, \eta)$-optimal solution of \cref{opt:genRA} or establishes that no such solution exists.
\end{theorem}
\begin{IEEEproof}
	Please refer to the appendix.
\end{IEEEproof}

\begin{algorithm}
	\caption{\cGls{sit} Algorithm for \cref{opt:genRA}}\label{alg:sitbb}
	\small
	\centering
	\begin{minipage}{\linewidth-1em}
	\begin{enumerate}[label=\textbf{Step \arabic*},ref=\arabic*,start=0,leftmargin=*]
		\item\label{alg:sitbb:0} Initialize $\varepsilon, \eta > 0$ and $\mathcal M_0 = [\vec p^0, \vec q^0]$ as in \cref{eq:M0}, 
			$\mathscr P_1 = \{ \mathcal M_0 \}$, $\mathscr R = \emptyset$, and $k = 1$.
			Initialize $\bar{\vec x}$ with the best known nonisolated feasible solution and set $\gamma$ as described in Step~\ref{alg:sitbb:4}; otherwise do not set $\bar{\vec x}$ and choose $\gamma \le \frac{f^+(\vec x, \vec\xi)}{f^-(\vec x, \vec\xi)}$ for all feasible $(\vec x, \vec\xi)$.
		\item\label{alg:sitbb:1} For each box $\mathcal M\in\mathscr P_k$:
			\begin{itemize}
				\item Compute $\beta(\mathcal M)$. Set $\beta(\mathcal M) = \infty$ if \cref{opt:bound} (or~\cref{opt:bound3}) is infeasible.
				\item Add $\mathcal M$ to $\mathscr R$ if $\beta(\mathcal M) \le -\varepsilon$.
			\end{itemize}
		\item\label{alg:sitbb:2} Terminate if $\mathscr R = \emptyset$: If $\bar{\vec x}$ is not set, then \cref{opt:genRA} is $\varepsilon$-essential infeasible; else $\bar{\vec x}$ is an essential $(\varepsilon, \eta)$-optimal solution of \cref{opt:genRA}.
		\item\label{alg:sitbb:3} Let $\mathcal M^k = \argmin\{\beta(\mathcal M) \,|\, \mathcal M\in\mathscr R\}$. Let $\vec x^k$ be the optimal solution of \cref{opt:bound} for the box $\mathcal M^k$ (or $\ubar{\vec x}^\ast_{\mathcal M^k}$ if \cref{opt:bound3} is employed for bounding), and $\vec y^k = \bar{\vec x}^\ast_{\mathcal M^k}$. Solve the feasibility problem
			\begin{fleqn}
			\begin{optprob}
				& \text{find}
				&& \vec\xi\in\mathcal C_{\vec x^k} \\
				& \text{s.\,t.}
				&& g_i^+(\vec x^k, \vec\xi) - g_i^-(\vec x^k) \le 0,\enskip \mathrlap{i = 1, \dots, m.}
				\label{opt:feas}
			\end{optprob}
			\end{fleqn}
			If \cref{opt:feas} is feasible go to Step~\ref{alg:sitbb:4}; otherwise go to Step~\ref{alg:sitbb:5}.
		\item\label{alg:sitbb:4} $\vec x^k$ is a nonisolated feasible solution satisfying $\frac{f^+(\vec x^k, \vec\xi)}{f^-(\vec x^k, \vec\xi)}\ge\gamma$ for some $\vec\xi\in\mathcal C_{\vec x^k}$. Let $\vec\xi^\ast$ be a solution to
			\begin{fleqn}
			\begin{optprob}
				& \underset{\vec\xi\in\mathcal C_{\vec x^k}}{\text{min}}
				&& \frac{f^+(\vec x^k, \vec\xi)}{f^-(\vec x^k, \vec\xi)} \\
				& \text{s.\,t.}
				&& g_i^+(\vec x^k, \vec\xi) - g_i^-(\vec x^k) \le 0,\quad \mathrlap{i = 1, 2 \dots, m.}
				\label{opt:sitbb}
			\end{optprob}
			\end{fleqn}
			If $\bar{\vec x}$ is not set or $\frac{f^+(\vec x^k, \vec\xi^\ast)}{f^-(\vec x^k, \vec\xi^\ast)} > \gamma-\eta$, set $\bar{\vec x} = \vec x^k$ and $\gamma = \frac{f^+(\vec x^k, \vec\xi^\ast)}{f^-(\vec x^k, \vec\xi^\ast)} + \eta$.
		\item\label{alg:sitbb:5} Bisect $\mathcal M^k$ via $(\vec v^k, j_k)$ where $j_k \in \argmax_j\{\abs{y^k_j - x^k_j}\}$ and $\vec v^k = \frac{1}{2} (\vec x^k + \vec y^k)$ (cf.~\cref{eq:bisect}).
			Remove $\mathcal M^k$ from $\mathscr R$. Let $\mathscr P_{k+1} = \{\mathcal M^k_-, \mathcal M^k_+\}$. Increment $k$ and go to Step~\ref{alg:sitbb:1}.
	\end{enumerate}
	\end{minipage}
\end{algorithm}

Observe that the \gls{bb} procedure is directly incorporated in \cref{alg:sit}. Thus, except for the first cycle, the \gls{bb} procedure is started from the boxes already in $\mathscr R$ instead of starting from scratch \cite[Remark~2]{Tuy2005a}. Also observe that $\mathscr R$ is a priority queue \cite[Sect.~5.2.3]{Knuth1997vol3}, and that \cref{opt:feas} is a convex optimization problem.
Due to the assumption of $\vec\xi$ being non-global variables, problem \cref{opt:sitbb} is efficiently solvable. For example,
if $f^+(\vec x, \vec\xi)$ is non-negative and concave in $\vec\xi$, and $f^-(\vec x, \vec \xi)$ is convex in $\vec\xi$ it is solvable with Dinkelbach's algorithm. Furthermore, if $f^-(\vec x, \vec \xi)$ is affine, $f^+(\vec x, \vec\xi)$ may also be negative. Please refer to \cite{Avriel1988,Schaible1993,Zappone2015} for more detailed treatments of fractional programming.

\begin{remark}
	The original algorithm in \cite{Tuy2016} contains an additional reduction step before computing the bound of a box. This step is optional and should only be included if it speeds up the algorithm. A direct application of the approach outlined in \cite{Tuy2016} requires several convex optimization problems to be solved for each reduction. In our numerical experiments this slowed down the algorithm and is thus omitted. We leave the design of an efficient reduction procedure open for future work.
\end{remark}

\subsection{Extension: Pointwise Minimum} \label{sec:ext:min}
\Cref{alg:sitbb} is easily extended to the case were the objective is the pointwise minimum of several functions, i.e.,
\begin{optprob}
	& \underset{(\vec x, \vec\xi) \in \mathcal C}{\text{max}} 
	&& \min_j\bigg\{ \frac{f_j^+(\vec x, \vec\xi)}{f_j^-(\vec x, \vec\xi)} \bigg\} \\
	& \text{s.\,t.}
	&& g_i^+(\vec x, \vec\xi) - g_i^-(\vec x) \le 0,\quad i = 1, 2, \dots, m.
\end{optprob}
In that case, all fractions in the minimum need to be greater or equal than $\gamma$ and the dual problem \cref{opt:dualRA} becomes
\begin{optprob}\label{opt:dualRAmin}
	& \underset{(\vec x, \vec\xi) \in \mathcal C}{\text{min}} 
	&& \underset{i = 1, 2, \dots, m}{\text{max}} \left( g_i^+(\vec x, \vec\xi) - g_i^-(\vec x) \right) \\
	& \text{s.\,t.}
	&& \gamma f_j^-(\vec x, \vec\xi) - f_j^+(\vec x, \vec\xi) \le 0,\quad j = 1, 2, \dots
\end{optprob}
Then, the branching and bounding procedures are easily adjusted to this extended dual problem \cref{opt:dualRAmin}.

\section{Numerical Evaluation} \label{sec:numeval}
The application of any optimization framework to a specific optimization problem requires, in general, some transformation of the initial problem to bring it into a form suitable for the framework. In addition, modification of the problem often allows to reduce the computational complexity significantly.

In this section, we first present the system model of a specific interference network, namely the Gaussian \cgls{mwrc}, and two of its achievable rate region. We then formulate the resource allocation problems and discuss, based on \cref{sec:RA}, the application of \cref{alg:sitbb}. Subsequently, we employ \cref{alg:sitbb} to obtain throughput and \cgls{gee} optimal resource allocations and compare the performance of \cref{alg:sitbb} to the state-of-the-art.
We conclude this section by evaluating how the performance of \cref{alg:sitbb} scales with an increasing number of global and non-global variables in \cref{sec:numeval:benchmark}.

\subsection{System Model}
We consider a 3-user \cgls{siso} Gaussian \cgls{mwrc} \cite{Gunduz2013,Chaaban2015} with \cgls{af} relaying, multiple unicast transmissions and no direct user-to-user links.
Users are indexed by $k$, $k\in\mathcal K=\{1, 2, 3\}$, and the relay is node 0.
User $k$ transmits with power $P_k$ over the channel $h_k$ to the relay. The relay propagates the observed symbol back to the users with transmit power $P_0$ over the channels $g_k$, $k\in\mathcal K$. Each node observes the \gls{iid} zero-mean circularly symmetric complex Gaussian noise with power $N_k$, $k\in\mathcal K\cup\{0\}$ and is subject to an average power constraint $\bar P_k$ on $X_k$, $k\in\mathcal K\cup\{0\}$.
The message exchange is defined by the functions $q: \mathcal K \mapsto \mathcal K$ and $l: \mathcal K \mapsto \mathcal K$, where the receiver of node $k$'s message is $q(k)$ and the user not interested in it is $l(k)$.
Without loss of generality we assume $q(1) = l(3) = 2$, $q(2) = l(1) = 3$, and $q(3) = l(2) = 1$. Due to space constraints we refer the reader to \cite{Matthiesen2015a,camsap17,spawc18} for a more detailed treatment of this system model.

The receiver uses \cgls{snd} which is the optimal decoder for interference networks when restricted to random codebooks with superposition coding and time sharing \cite{bandemer2015}. We consider two different codebook constructions: traditional single message encoding and Han-Kobayashi \cite{Han1981} inspired rate splitting.

\subsubsection{Single Message}
The achievable rate region is given below where $S_k = \frac{P_k}{N_0}$, $\vec S = (S_k)_{k\in\mathcal K}$, and $\bar{\vec S} = (\bar S_k)_{k\in\mathcal K}$. Observe that this region is strictly larger than previously published \cgls{snd} regions \cite{Matthiesen2015a,camsap17} and includes \cgls{ian} as a special case. This is due to recent insights on \cgls{snd} decoders \cite{bandemer2015}.
\begin{lemma} \label{lem:snd}
	A rate triple $(R_1, R_2, R_3)$ is achievable for the Gaussian MWRC with AF and \gls{snd} if, for each $k\in\mathcal K$,
	\begin{equation} \label{lem:snd:ian}
		R_k \le \log\left( 1 + \frac{\abs{h_k}^2 S_k}{\gamma_{k}(\vec S)} \right)
	\end{equation}
	or
	\begin{subequations} \label{lem:snd:snd}
	\begin{align}
		R_k &\le \log\left( 1 + \frac{\abs{h_k}^2 S_k}{\delta_{k}(\vec S)} \right) \\
		R_k + R_{l(k)} &\le \log\left( 1 + \frac{\abs{h_k}^2 S_k + \abs{h_{l(k)}}^2 S_{l(k)}}{\delta_{k}(\vec S)} \right)
	\end{align}
	\end{subequations}
	where
	$S_k  \le \bar S_k$, $\delta_k(\vec{S}) = 1 + \widetilde{g}\qk^{-1} \left( 1 + \sum_{i\in\mathcal K} \abs{h_i}^2 S_i \right)$
	with $\widetilde{g}_k = \abs{g_k}^2 \frac{\bar P_0}{N_k}$, and $\gamma_k(\vec S) = \delta_k(\vec S) + \abs{h\lk}^2 S\lk$.
\end{lemma}

\begin{IEEEproof}
	The proof follows along the lines of \cite[Sect.~II-A.]{bandemer2015} and is omitted due to space constraints.
\end{IEEEproof}

Let $\mathcal R_{k,\text{IAN}}$ and $\mathcal R_{k,\text{SND}}$ be the regions defined by \cref{lem:snd:ian} and \cref{lem:snd:snd}, respectively. Then, the rate region in \cref{lem:snd} is
\begin{equation*}
	\mathcal R = \bigcap_{k\in\mathcal K} (\mathcal R_{k,\text{IAN}} \cup \mathcal R_{k,\text{SND}}) = \bigcup_{\vec d\in\{\text{IAN}, \text{SND}\}^{\card{\mathcal K}}} \bigcap_{k\in\mathcal K} \mathcal R_{k,d_k}.
\end{equation*}
Since
$\inf_{\vec x\in\bigcup_i \mathcal D_i} f(\vec x) = \min_i \inf_{\vec x\in\mathcal D_i} f(\vec x)$, we can split the resource allocation problem for \cref{lem:snd} into eight individual optimization problems. Each is easily identified as an instance of \cref{opt:exRA} and solvable with \cref{alg:sitbb} using the initial box $\mathcal M_0 = [\vec 0, \bar{\vec S}]$.

\subsubsection{Rate Splitting}
Each message is divided into a common message to be decoded by all receivers and a private part that is treated as additional noise by unconcerned receivers.
These messages are then encoded by individual Gaussian codebooks with powers $P_k^c$ and $P_k^p$ and linearly superposed to be transmitted in a single codeword with power $P_k = P_k^c + P_k^p$.
The achievable rate region is given below in terms of the \cglspl{snr} $S_k^c = \frac{P_k^c}{N_0}$ and $S_k^p = \frac{P_k^p}{N_0}$. Further, we define 
$S_k = S_k^c + S_k^p$, $\bar S_k = \frac{\bar P_k}{N_0}$, $\vec S^c = (S_k^c)_{k\in\mathcal K}$, $\vec S^p = (S_k^p)_{k\in\mathcal K}$, $\vec S = (\vec S^c, \vec S^p)$, and $\bar{\vec S} = (\bar S_k)_{k\in\mathcal K}$.
\begin{lemma} \label{lem:HK}
	A rate triple $(R_1, R_2, R_3)$ is achievable for the Gaussian \cgls{mwrc} with \cgls{af} relaying if, for all $k\in\mathcal K$,
\begingroup
\belowdisplayskip=0pt
	\begin{subequations}
	\begin{flalign}
		&& R_k &\le \BiShort{k}{q(k)}{l(k)}, \label{eq:HK1}\\
		&& R_k + R\qk &\le \AiShort{k}{q(k)}{l(k)} + \DiShort{q(k)}{l(k)}{k}, \label{eq:HK2}\\
		&& R_{k} + R_{q(k)} + R_{l(k)} &\le \AiShort{k}{q(k)}{l(k)} +  \CiShort{q(k)}{l(k)}{k} + \DiShort{l(k)}{k}{q(k)}, \label{eq:HK3}\\
		&& 2 R_k + R_{q(k)} + R_{l(k)} &\le \AiShort{k}{q(k)}{l(k)} + \CiShort{q(k)}{l(k)}{k} + \CiShort{l(k)}{k}{q(k)} + \DiShort{k}{q(k)}{l(k)},\label{eq:HK4}\\
		%
		%
		\text{\normalsize and,} && R_1 + R_2 + R_3 &\le \CiShort{1}{2}{3} + \CiShort{2}{3}{1} + \CiShort{3}{1}{2}, \label{eq:HK5}& 
	\end{flalign}%
	\end{subequations}
\endgroup
\begingroup
\abovedisplayskip=0pt
	\begin{subequations}\label{eq:defABCD}
	\begin{flalign}
		\text{\normalsize with } && \AiShort{k}{q(k)}{l(k)} &= \Ai{k}{q(k)}{l(k)} \label{eq:defAk} \\
		&& \BiShort{k}{q(k)}{l(k)} &= \Bi{k}{q(k)}{l(k)} \label{eq:defBk} \\
		&& \CiShort{k}{q(k)}{l(k)} &= \Ci{k}{q(k)}{l(k)} \label{eq:defCk} \\
		&& \DiShort{k}{q(k)}{l(k)} &= \Di{k}{q(k)}{l(k)} \label{eq:defDk}  
	\end{flalign}
	\end{subequations}
\endgroup
	where
	$S^c_k + S^p_k  \le \bar S_k$ and
\begingroup
\abovedisplayskip=1pt
\belowdisplayskip=1pt
	\begin{equation*}
		\small
		\gamma_k(\vec{S}) = 1 + \abs{h\lk}^2 S^p\lk + \widetilde{g}\qk^{-1} \left( 1 + \sum_{i\in\mathcal K} \abs{h_i}^2 (S^c_i + S^p_i) \right) ,
	\end{equation*}
\endgroup
	with $\widetilde{g}_k = \abs{g_k}^2 \frac{\bar P_0}{N_k}$.
\end{lemma}
\begin{IEEEproof}
	Please refer to the appendix.
\end{IEEEproof}

Global optimal resource allocation for this scenario is a straightforward extension of \cref{opt:exRA}:
\begin{optprob} \label{opt:HK}
	& \underset{\vec S, \vec R}{\text{max}} 
	&& f(\vec S, \vec R) \\
	& \text{s.\,t.}
	&& \vec a_i^T \vec R \le \sum_j \log\!\left( 1 + \frac{\vec b_{i,j}^T \vec S}{\gamma_{\kappa(i,j)}(\vec S)} \right),\quad i = 1, \ldots, n \\
	&&& S_k^c + S_k^p \le \bar S_k,\enskip k\in\mathcal K,\qquad \vec R \ge 0,\quad \vec S \ge 0
\end{optprob}
where $\kappa(i,j)$ maps from $(i, j)$ to the correct $k\in\mathcal K$ and $\vec a_i, \vec b_{i,j} \ge 0$ are easily identified from \cref{lem:HK}. From the discussion in \cref{sec:RA} it is apparent that \cref{opt:HK}, depending on the identification of $g^+_i(\vec S, \vec R)$ and $g^-_i(\vec S)$, satisfies both, \cref{constrA,constrB}, and is solvable with \cref{alg:sitbb} where the global and non-global variables are $\vec S$ and $\vec R$, respectively.
However, we can reduce the number of global variables in \cref{opt:HK} from six to four: When using the identification $g^-_i(\vec S)\coloneqq -\sum_j \log( \gamma_{\kappa(i,j)}(\vec S) )$, the non-convexity due to $\vec S^c$ stems only from the sum $\sum_{k\in\mathcal K} \abs{h_k}^2 S^c_k$ in $\gamma_k(\vec S)$. Thus, if we replace this sum by an auxiliary variable $y$, the variables $\vec S^c$ can be treated as non-global. The resulting problem is
\begin{optprob} \label{opt:HK2}
	& \underset{\vec S, \vec R, y}{\text{max}} 
	&& f(\vec S, \vec R) \\
	& \text{s.\,t.}
	&& \vec a_i^T \vec R \le \sum_j \log\!\left( 1 + \frac{\vec b_{i,j}^T \vec S}{\gamma_{\kappa(i,j)}(\vec S^p, y)} \right),\quad i = 1, \ldots, n \\
	&&& y \ge \sum_{k\in\mathcal K} \abs{h_k}^2 S^c_k \\
	&&& S_k^c + S_k^p \le \bar S_k,\enskip k\in\mathcal K,\qquad \vec R \ge 0,\quad \vec S \ge 0
\end{optprob}
where $(\vec S^p, y)$ are the global and $(\vec S^c, \vec R)$ are the non-global variables.
Since the computational complexity grows exponentially in the number of global variables, solving \cref{opt:HK2} instead of \cref{opt:HK} reduces the complexity significantly. The drawback of this approach is that \cref{constrB} is no longer satisfied and the bounding problem becomes convex instead of linear. However, the performance gain outweighs these, comparatively small, performances losses significantly. The proposition below formally states the equivalence of \cref{opt:HK,opt:HK2}.

\begin{proposition} \label{prop:equiv}
	Problems~\cref{opt:HK,opt:HK2} are equivalent in the sense that $v\eqref{opt:HK} = v\eqref{opt:HK2}$ if $f(\vec S, \vec R)$ is increasing in $\vec R$ for fixed $\vec S$.
\end{proposition}
\begin{IEEEproof}
	Please refer to the appendix.
\end{IEEEproof}

Finally, note that all constraints in \cref{opt:HK2} except the first are linear and, thus, $\mathcal C$ is a convex set. The initial box $\mathcal M_0$ required by \cref{alg:sitbb} is easily identified as $[\vec 0, \bar{\vec S}] \times [0, \sum_{k\in\mathcal K} \abs{h_k}^2 \bar S_k]$.

\subsection{Throughput}
We employ \cref{alg:sitbb} to compute the throughput optimal resource allocation for the rate regions in \cref{lem:snd,lem:HK}, i.e., the objective is $f(\vec S, \vec R) = \vec w^T \vec R$ with $w_i = 1$ for all $i$ and obviously fulfills \cref{prop:equiv}.
We assume equal maximum power constraints and noise power at the users and the relay, and no minimum rate constraint, i.e. $\ubar{\vec R} = \vec 0$. Channels are assumed reciprocal and chosen \gls{iid} with circular symmetric complex Gaussian distribution, i.e., $h_k \sim \mathcal{CN}(0, 1)$ and $g_k = h_k^*$. Results are averaged over 1,000 channel realizations. The precision of the objective value is $\eta = 0.01$, $\varepsilon$ is chosen as $10^{-5}$\!, and the algorithm is started with $\gamma = 0$.

\begin{figure}
\centering
\begin{tikzpicture}
	\begin{axis} [
			thick,
			xlabel={SNR [dB]},
			ylabel={Sum Rate [bit/s/Hz]},
			ylabel near ticks,
			grid=major,
			xtick = {0,5,...,30},
			ytick = {0, 2.5, 5, ..., 20},
			minor y tick num = 1,
			yminorgrids = true,
			mark repeat = 1,
			xmin = 0,
			xmax = 25,
			ymin = 0,
			legend pos=north west,
			legend cell align=left,
			cycle list name=default,
		]

		\pgfplotstableread[col sep=comma]{sumrate.dat}\tbl

		\addplot+[smooth] table[y=HK] {\tbl};
		\addlegendentry{RS};

		\addplot+[smooth] table[y=SND] {\tbl};
		\addlegendentry{SND};

		\addplot+[smooth] table[y=Joint] {\tbl};
		\addlegendentry{``traditional'' SND};

		\addplot+[smooth] table[y=TIN] {\tbl};
		\addlegendentry{IAN};
	\end{axis}
\end{tikzpicture}
\vspace{-2ex}
\caption{Throughput in the MWRC with AF relaying and 1) \cgls{rs} 2) \cgls{snd}; 3) ``traditional'' \cgls{snd}; and 4) \cgls{ian}. Averaged over 1000 \cgls{iid} channel realizations.}
\label{fig:sr}
\vspace{-2ex}
\end{figure}
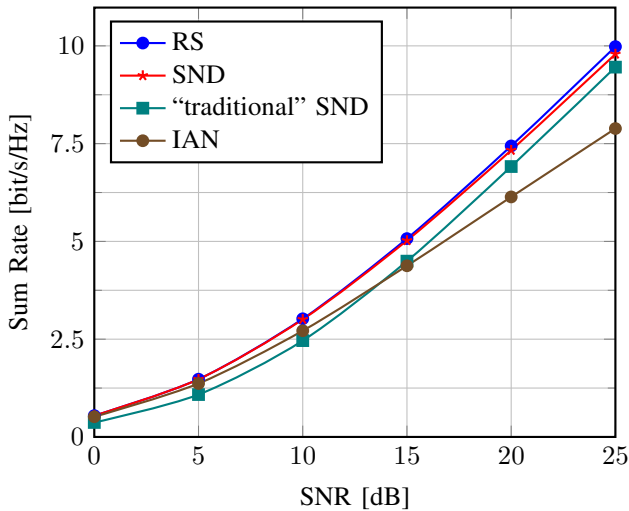

\Cref{fig:sr} displays the maximum throughput for \cgls{snd} as in \cref{lem:snd} and \cgls{rs} as in \cref{lem:HK}. Before \cite{bandemer2015}, common wisdom was that neither \cgls{ian} nor \cgls{snd} dominates the other rate-wise, with \cgls{ian} generally better in noise limited scenarios and \cgls{snd} superior when interference is the limiting factor. This misconception is due to an longstanding oversight in the \cgls{snd} proof that was clarified in \cite{bandemer2015}. \Cref{fig:sr} also includes results for ``traditional'' \cgls{snd} and \cgls{ian} as obtained rate regions defined by \cref{lem:snd:snd,lem:snd:ian}, respectively.
First, observe that, in accordance with conventional wisdom, neither ``traditional'' \cgls{snd} nor \cgls{ian} dominates the other. Instead, ``extended'' \cgls{snd} clearly dominates the other two where the gain is solely due to allowing each receiver to either use \cgls{ian} or ``traditional'' \cgls{snd}. The average gain of \cgls{snd} over the other two is approximately \unit[11]{\%} and \unit[22]{\%} at \unit[10]{dB}, respectively, or \unit[0.29]{bpcu} and \unit[0.54]{bpcu}. Note that this gain is only achieved by allowing each receiver to choose between \cgls{ian} and joint decoding which does not result in higher decoding complexity than ``traditional'' \cgls{snd}.
The average gain observed for \cgls{rs} over single message \cgls{snd} is rather small, e.g., at \unit[25]{dB} it is only \unit[2]{\%} (or \unit[0.2]{bpcu}). However, depending on the channel realization we observed gains up to half a bit (or \unit[4.6]{\%}) at \unit[20]{dB}. With spectrum being an increasingly scarce
and expensive
resource this occasional gain might
very well
justify the slightly higher coding complexity.

\tikzsetnextfilename{SR}
\tikzpicturedependsonfile{sumrate.dat}

A popular approach to solving non-convex resource allocation problems with global optimality is to use monotonic optimization \cite{Bjornson2013,Zappone2017}. As already pointed out in the introduction, the main challenge is that for \cgls{snd} the optimization is over the rates and powers (instead of just the powers for \cgls{ian}). Thus, every approach that treats the rates as global variables will suffer from very long computation time. Instead, in \cite{camsap17} the problem is solved by decomposing it into an inner linear and an outer monotonic program. We use this approach as the state-of-the-art reference for our performance comparison. The inner linear problem is solved with Gurobi~8 \cite{gurobi8} and the monotonic program with the Polyblock algorithm\footnote{The Polyblock algorithm was implemented in C++ with similar techniques as in the implementation of \cref{alg:sitbb}. Thus, performance differences should be mostly due to algorithmic differences.} \cite{Tuy2000} with a tolerance of 0.01. Results match those obtained with our \cgls{sit} algorithm for ``traditional'' \cgls{snd} and, thus, are not displayed in \cref{fig:sr}. Run times for both algorithms are reported in \cref{tab:srrt}. About \unit[5.4]{\%} of the computations with the Polyblock algorithm did not complete within one week. For these, a run time of one week was assumed for the computation of the mean in \cref{tab:srrt}. Thus, the reported run times for the Polyblock algorithm are an underestimate. Nevertheless, we observe, on average, roughly 10,000$\times$ faster convergence for the \cgls{sit} algorithm for low to medium \cglspl{snr}. Interestingly, the median run time for high \cglspl{snr} is lower for the Polyblock algorithm while the \cgls{sit} is clearly faster on average.

\begin{table}
	\pgfplotstableread[col sep=comma]{sumrate_runtime.dat}\tbl
	\pgfplotstableforeachcolumnelement{snr}\of\tbl\as\cell{%
		\pgfmathequal{0}{\cell}
		\ifnum\pgfmathresult>0
			\let\zeroDB=\pgfplotstablerow
		\fi
		\pgfmathequal{15}{\cell}
		\ifnum\pgfmathresult>0
			\let\fifteenDB=\pgfplotstablerow
		\fi
		\pgfmathequal{30}{\cell}
		\ifnum\pgfmathresult>0
			\let\thirtyDB=\pgfplotstablerow
		\fi
	}%
	\newlength{\mylen}
	\settowidth{\mylen}{``trad.''}
	%
	%
	\centering
	\caption{Mean and median run times of throughput maximization for \cgls{snd} and ``traditional'' \cgls{snd}. For ``traditional'' \cgls{snd}, both the \cgls{sit} and the Polyblock algorithm (PA) \cite{camsap17} are employed.}
	\label{tab:srrt}
	\begin{tabular}{rrrrr}
		\toprule
		& & \multicolumn{3}{c}{SNR} \\
		& & \unit[0]{dB} & \unit[15]{dB} & \unit[30]{dB} \\
		\midrule
		\multirow{2}{*}{SND} & Mean
			& \unit[\pgfplotstablegetelem{\zeroDB}{SND_mean}\of\tbl\pgfmathprintnumber{\pgfplotsretval}]{s}
			& \unit[\pgfplotstablegetelem{\fifteenDB}{SND_mean}\of\tbl\pgfmathprintnumber{\pgfplotsretval}]{s}
			& \unit[\pgfplotstablegetelem{\thirtyDB}{SND_mean}\of\tbl\pgfmathprintnumber{\pgfplotsretval}]{s} \\
		& Median 
			& \unit[\pgfplotstablegetelem{\zeroDB}{SND_median}\of\tbl\pgfmathprintnumber{\pgfplotsretval}]{s}
			& \unit[\pgfplotstablegetelem{\fifteenDB}{SND_median}\of\tbl\pgfmathprintnumber{\pgfplotsretval}]{s}
			& \unit[\pgfplotstablegetelem{\thirtyDB}{SND_median}\of\tbl\pgfmathprintnumber{\pgfplotsretval}]{s} \\
		\midrule
		\multirow{2}{\mylen}{\raggedleft``trad.'' SND} & Mean
			& \unit[\pgfplotstablegetelem{\zeroDB}{Joint_mean}\of\tbl\pgfmathprintnumber{\pgfplotsretval}]{s}
			& \unit[\pgfplotstablegetelem{\fifteenDB}{Joint_mean}\of\tbl\pgfmathprintnumber{\pgfplotsretval}]{s}
			& \unit[\pgfplotstablegetelem{\thirtyDB}{Joint_mean}\of\tbl\pgfmathprintnumber{\pgfplotsretval}]{s} \\
		& Median 
			& \unit[\pgfplotstablegetelem{\zeroDB}{Joint_median}\of\tbl\pgfmathprintnumber{\pgfplotsretval}]{s}
			& \unit[\pgfplotstablegetelem{\fifteenDB}{Joint_median}\of\tbl\pgfmathprintnumber{\pgfplotsretval}]{s}
			& \unit[\pgfplotstablegetelem{\thirtyDB}{Joint_median}\of\tbl\pgfmathprintnumber{\pgfplotsretval}]{s} \\
		\midrule
		\multirow{2}{*}{PA} & Mean
			& \unit[\pgfplotstablegetelem{\zeroDB}{PA_mean}\of\tbl\pgfmathprintnumber{\pgfplotsretval}]{s}
			& \unit[\pgfplotstablegetelem{\fifteenDB}{PA_mean}\of\tbl\pgfmathprintnumber{\pgfplotsretval}]{s}
			& \unit[\pgfplotstablegetelem{\thirtyDB}{PA_mean}\of\tbl\pgfmathprintnumber{\pgfplotsretval}]{s} \\
		& Median 
			& \unit[\pgfplotstablegetelem{\zeroDB}{PA_median}\of\tbl\pgfmathprintnumber{\pgfplotsretval}]{s}
			& \unit[\pgfplotstablegetelem{\fifteenDB}{PA_median}\of\tbl\pgfmathprintnumber{\pgfplotsretval}]{s}
			& \unit[\pgfplotstablegetelem{\thirtyDB}{PA_median}\of\tbl\pgfmathprintnumber{\pgfplotsretval}]{s} \\
		\bottomrule
	\end{tabular}
\end{table}

\subsection{Energy Efficiency}
\begin{figure}
\centering
\begin{tikzpicture}
	\begin{axis} [
			thick,
			xlabel={SNR [dB]},
			ylabel={EE [bit/Joule/Hz]},
			ylabel near ticks,
			grid=major,
			ytick = {0, .1, .2, .3},
			minor y tick num = 3,
			yminorgrids = true,
			mark repeat = 2,
			xmin = -15,
			xmax = 40,
			ymin = 0,
			legend pos=south east,
			legend cell align=left,
			cycle list name=default,
		]

		\pgfplotstableread[col sep=comma]{ee.dat}\tbl

		\addplot+[smooth] table[y=SND] {\tbl};
		\addlegendentry{SND};

		\addplot+[smooth] table[y=Joint] {\tbl};
		\addlegendentry{``traditional'' SND};

		\addplot+[smooth, mark phase = 2] table[y=TIN] {\tbl};
		\addlegendentry{IAN};
	\end{axis}
\end{tikzpicture}
\vspace{-2ex}
\caption{Energy efficiency in the MWRC with AF relaying and 1) \cgls{snd}; 2) ``traditional'' \cgls{snd}; and 3) \cgls{ian}. Averaged over 1000 \cgls{iid} channel realizations and computed with a precision of $\eta = 10^{-3}$.}
\label{fig:ee}
\vspace{-2ex}
\end{figure}
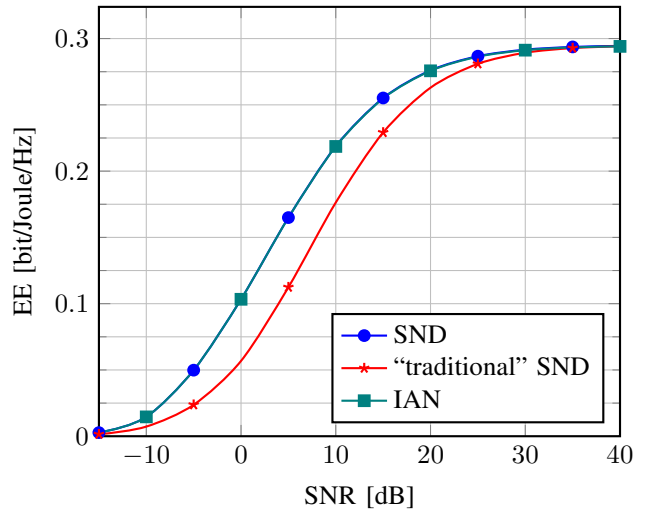

The \cgls{ee} as defined in \cref{eq:gee} is computed with the same parameters as before except for the precision which is chosen as $\eta = 10^{-3}$ for \cref{fig:ee} and $\eta = 10^{-2}$ for \cref{tab:eert}. Additionally, we assume the static circuit power $P_c = \unit[1]{W}$, the power amplifier inefficiencies $\phi_i = 4$, and that the relay always transmits at maximum power. Results for \cgls{snd}, ``traditional'' \cgls{snd}, and \cgls{ian} are displayed in \cref{fig:ee}. First, observe that the curves saturate starting from \unit[30]{dB} as is common for \cgls{ee} maximization. In this saturation region, all three approaches achieve the same \cgls{ee}. However, for lower \cglspl{snr}, \cgls{ian} and, thus also \cgls{snd}, outperform ``traditional'' \cgls{snd} by \unit[24]{\%} on average at \unit[10]{dB}. Of course, the \cgls{ee} performance depends quite a lot on the choice of real-world simulation parameters \cite{Matthiesen2015,Bjornson2017}, so further work is necessary to draw final conclusions in this regard.

\tikzsetnextfilename{EE}
\tikzpicturedependsonfile{ee.dat}

The main point of this subsection, however, is the performance comparison of different computational approaches to the \cgls{ee} computation. Recall from \cref{sec:RA} that there are two possible identifications of $g_i^+$ and $g_i^-$ that result in the problem either belonging to \cref{constrA} or \cref{constrB}. As discussed before, identification \cref{eq:identification1} (resp.~\cref{constrA}) leads to tighter bounds than \cref{eq:identification2} but requires the solution of a convex optimization problem in the bounding step. Instead, \cref{eq:identification2} (resp.~\cref{constrB}) leads to a linear bounding problem which is considerably easier to solve. \Cref{tab:eert} summarizes mean and median computations times for both approaches. Each bounding problem is solved with the fastest solver available: the convex problem that stems from \Cref{constrA} with Mosek \cite{mosek} and the linear from \cref{constrB} with Gurobi \cite{gurobi8}. Observe, that Gurobi is, on average, thrice as fast as Mosek at \unit[0]{dB} and Mosek is almost five times faster than Gurobi at \unit[40]{dB}. However, these are only relative numbers. The average total computation time per channel realization is \unit[8]{s} for Gurobi and \unit[19]{s} for Mosek. So, despite the faster convergence speed of \cref{constrA} (i.e., Mosek), \cref{constrB} is much faster due to the lower computational complexity of the linear bounding problem.

\begin{table}
	\pgfplotstableread[col sep=comma]{ee_runtime.dat}\tbl
	\pgfplotstableforeachcolumnelement{snr}\of\tbl\as\cell{%
		\pgfmathequal{20}{\cell}
		\ifnum\pgfmathresult>0
			\let\twentyDB=\pgfplotstablerow
		\fi
		\pgfmathequal{0}{\cell}
		\ifnum\pgfmathresult>0
			\let\zeroDB=\pgfplotstablerow
		\fi
		\pgfmathequal{40}{\cell}
		\ifnum\pgfmathresult>0
			\let\fortyDB=\pgfplotstablerow
		\fi
	}%
	\tikzset{/pgf/number format/precision=4}
	\centering
	\caption{Mean and median run times of \cgls{ee} computation for ``traditional'' \cgls{snd} and different solvers, all with precision $\eta = 0.01$}
	\label{tab:eert}
	\begin{tabular}{rrrrr}
		\toprule
		& & \multicolumn{3}{c}{SNR} \\
		& & \unit[0]{dB} & \unit[20]{dB} & \unit[40]{dB} \\
		\midrule
		\multirow{2}{*}{Gurobi} & Mean
			& \unit[\pgfplotstablegetelem{\zeroDB}{Joint_gurobi_mean}\of\tbl\pgfmathprintnumber{\pgfplotsretval}]{s}
			& \unit[\pgfplotstablegetelem{\twentyDB}{Joint_gurobi_mean}\of\tbl\pgfmathprintnumber{\pgfplotsretval}]{s}
			& \unit[\pgfplotstablegetelem{\fortyDB}{Joint_gurobi_mean}\of\tbl\pgfmathprintnumber{\pgfplotsretval}]{s} \\
		& Median 
			& \unit[\pgfplotstablegetelem{\zeroDB}{Joint_gurobi_median}\of\tbl\pgfmathprintnumber{\pgfplotsretval}]{s}
			& \unit[\pgfplotstablegetelem{\twentyDB}{Joint_gurobi_median}\of\tbl\pgfmathprintnumber{\pgfplotsretval}]{s}
			& \unit[\pgfplotstablegetelem{\fortyDB}{Joint_gurobi_median}\of\tbl\pgfmathprintnumber{\pgfplotsretval}]{s} \\
		\midrule
		\multirow{2}{*}{Mosek} & Mean
			& \unit[\pgfplotstablegetelem{\zeroDB}{Joint_mosek_mean}\of\tbl\pgfmathprintnumber{\pgfplotsretval}]{s}
			& \unit[\pgfplotstablegetelem{\twentyDB}{Joint_mosek_mean}\of\tbl\pgfmathprintnumber{\pgfplotsretval}]{s}
			& \unit[\pgfplotstablegetelem{\fortyDB}{Joint_mosek_mean}\of\tbl\pgfmathprintnumber{\pgfplotsretval}]{s} \\
		& Median
			& \unit[\pgfplotstablegetelem{\zeroDB}{Joint_mosek_median}\of\tbl\pgfmathprintnumber{\pgfplotsretval}]{s}
			& \unit[\pgfplotstablegetelem{\twentyDB}{Joint_mosek_median}\of\tbl\pgfmathprintnumber{\pgfplotsretval}]{s}
			& \unit[\pgfplotstablegetelem{\fortyDB}{Joint_mosek_median}\of\tbl\pgfmathprintnumber{\pgfplotsretval}]{s} \\
		\midrule
		\multirow{2}{*}{Dinkelbach} & Mean
			& \unit[\pgfplotstablegetelem{\zeroDB}{Dinkelbach_mean}\of\tbl\pgfmathprintnumber{\pgfplotsretval}]{s}
			& \unit[\pgfplotstablegetelem{\twentyDB}{Dinkelbach_mean}\of\tbl\pgfmathprintnumber{\pgfplotsretval}]{s}
			& \unit[\pgfplotstablegetelem{\fortyDB}{Dinkelbach_mean}\of\tbl\pgfmathprintnumber{\pgfplotsretval}]{s} \\
		& Median
			& \unit[\pgfplotstablegetelem{\zeroDB}{Dinkelbach_median}\of\tbl\pgfmathprintnumber{\pgfplotsretval}]{s}
			& \unit[\pgfplotstablegetelem{\twentyDB}{Dinkelbach_median}\of\tbl\pgfmathprintnumber{\pgfplotsretval}]{s}
			& \unit[\pgfplotstablegetelem{\fortyDB}{Dinkelbach_median}\of\tbl\pgfmathprintnumber{\pgfplotsretval}]{s} \\
		\bottomrule
	\end{tabular}
\end{table}

The state-of-the-art approach to compute the \cgls{ee} is Dinkelbach's Algorithm \cite{dinkelbach1967,Zappone2015}. This requires the global solution of a sequence of auxiliary problems with very similar properties to the throughput maximization problem in the previous section. Here, we solve this inner problem with the fastest method available: the \cgls{sit} algorithm developed in this paper (cf.~\cref{tab:srrt}). Hence, the differences in the run time are solely due to the use of Dinkelbach's Algorithm. It can be observed from \cref{tab:eert} that the inherent treatment of fractional objectives in our algorithm is always significantly faster (up to $800\times$ on average at \unit[20]{dB}) than Dinkelbach's Algorithm. Moreover, the obtained result is guaranteed to lie within an $\eta$-region around the true essential optimal value.

\subsection{Benchmark of \cref{alg:sitbb}} \label{sec:numeval:benchmark}
Above, we evaluated the performance of \cref{alg:sitbb} against the state-of-the-art on a real-world resource allocation problem. The goal of this subsection is to assess how \cref{alg:sitbb} scales with an increasing number of global and non-global variables.  This is done by means of an idealized example. Namely, consider a $K$-user \cgls{gic} with input-output relation
\begin{equation} \label{eq:ifc}
	y_k = \sum_{j=1}^K h_{kj} x_j + z_k
\end{equation}
where $h_{kj}\in\mathds C$ is the (effective) channel gain from transmitter $j$ to receiver $k$, $x_j$ is the complex-valued channel input of transmitter $j$ that is subject to an average power constraint $\bar P_j$, $y_k$ is the received symbol at receiver $k$, and $z_k$ is circularly-symmetric complex Gaussian noise with zero mean and power $N_k$. Consider just the $k$-th receiver which is interested in decoding $x_k$ and assume it jointly decodes a subset $\mathcal S_k\subseteq\{1, 2, \ldots, K\}$ of messages and treats the remaining $x_j$, $j\in\mathcal S_k^c = \{1, 2, \ldots, K\}\setminus\mathcal S_k$, as noise. This is a \cgls{mac} with capacity region
\begin{multline} \label{eq:RRmac}
	\mathcal A_k(\mathcal S_k) = \Bigg\{ (R_j)_{j\in\mathcal S_k} \,\Bigg|\, \sum_{i \in\mathcal T} R_i \le \log\left( 1+ \frac{P_k(\mathcal T)}{N_k + P_k(\mathcal S_k^c)} \right)\\ \text{ for every } \mathcal T\subseteq\mathcal S_k \rlap{\Bigg\}}\,
\end{multline}
where $P_k(\mathcal S) =  \sum_{j\in\mathcal S} \abs{h_{kj}}^2 P_j$. The achievable rate region for the complete system then is
\begin{equation}
	\mathcal R = \bigcap_{k=1}^K \mathcal A_k(\mathcal S_k).
\end{equation}

For the benchmark, assume that receiver $i$ jointly decodes its own message and the interfering message of transmitter $i+1 \mod K$. All other messages are treated as noise. Moreover, assume that $h_{ij} = 0$ for $i \neq j$, $j \neq i+1$, $j > \kappa$, and some positive $\kappa$. These assumptions allow to precisely control the complexities in the global and non-global variables. Thus, $\mathcal S_k = \{ k, k+1 \mod K \}$ for all $k$ and $\mathcal S_k^c = \{ 1, \dots, \kappa \}\setminus \mathcal S_k$.
Maximizing the throughput in this system requires the solution of
\begin{optprob} \label{opt:gic}
	& \underset{\vec p, \vec R}{\text{max}} 
	&& \sum_{k=1}^K R_k \\
	& \text{s.\,t.}
	&& \vec R \in \mathcal A_k(\mathcal S_k),\enskip \text{for all}\ k\\
	&&& \vec R \ge 0,\quad 0\le\vec p\le \bar{\vec P}.
\end{optprob}
First, observe that this is a non-convex optimization problem due to $P_k(\mathcal S_k^c)$ in the denominators of the $\log$-terms in \cref{eq:RRmac}. Further, observe that $\mathcal S_k^c \subseteq \{ 1, \dots, \kappa \}$ for all $k = 1, \dots, K$. Hence, we can identify the global variables as $(p_k)_{k=1}^\kappa$ and the non-global variables as $(p_k)_{\kappa+1}^K$ and $\vec R$. Moreover, the powers $(p_k)_{\kappa+1}^K$ only occur in the numerators of the $\log$-terms in \cref{eq:RRmac} and these $\log$-terms are increasing functions in $(p_k)_{\kappa+1}^K$. Thus, $p_k = \bar P_k$ is the optimal solution for all $k = \kappa+1, \dots, K$. Plugging-in this partial solutions in \cref{opt:gic}, we obtain a simplified version of \cref{opt:gic} with $\kappa$ global variables $(p_k)_{k=1}^\kappa$, $K$ linear variables $\vec R$, and $3 K$ inequality constraints. Thus, we can indepently control the number of global and non-global variables. Since the number of constraints grows linearly in the number of global variables and the bounding problem is a linear program, the computational complexity of solving \cref{opt:gic} should grow polynomially in the number of non-global variables.

This rationale is verified numerically in \cref{fig:benchUE} where \cref{opt:gic} is solved for a fixed number of $\kappa$ global variables and an increasing number of $K$ total users in the system.
We employ \cref{alg:sitbb} with Gurobi as inner solver and choose $\bar P_k = 1$ and $N_k = 0.01$ for all $k = 1, \dots, K$, i.e., a transmit \cgls{snr} of \unit[20]{dB}, and \cgls{iid} $h_{kj} \sim \mathcal{CN}(0,1)$. Each data point is obtained by averaging over 1,000 independent channel realizations. \Cref{alg:sitbb} was started with parameters $\eta = 0.01$, $\varepsilon = 10^{-5}$, and $\gamma = 0$.
Since the abcissa is directly proportional to the number of non-global variables ($K-\kappa$) and the number of inequality constraints ($3K$), we expect to observe a polynomial growth of the run time. Indeed, \cref{fig:benchUE} displays a roughly linear growth expect for small $K$. This is because, for big $K$, the optimal power allocation converges to setting all global variables to zero. Instead, for small $K$ these powers are generally non-zero and the global part of the optimization problem requires more time to converge to the optimal solution.

\begin{figure}
	\centering
	\tikzexternaldisable
	\tikzsetnextfilename{benchNC}
	\tikzpicturedependsonfile{benchNC.dat}
	\begin{tikzpicture}
		\begin{semilogyaxis} [
				thick,
				xlabel={Number of global variables $\kappa$},
				ylabel={Run Time [s]},
				ylabel near ticks,
				xlabel near ticks,
				grid,
				ymax = 1e4,
				ymin = 1e-2,
				xmax = 5,
				xmin = 2,
				xtick distance = 1,
				ytick distance = 10^1,
				mark repeat = 1,
				legend pos=south east,
				legend cell align=left,
				cycle list name=default,
			]

			\pgfplotstableread[col sep=comma]{benchNC.dat}\tbl

			\addplot table[y=UE7] {\tbl};
		\end{semilogyaxis}
	\end{tikzpicture}
	\vspace{-2ex}
	\caption{Average run time of \cref{alg:sitbb} to solve \cref{opt:gic} as a function of the number of global variables $\kappa$ for fixed number of $K = 7$ users. Averaged over 1,000 \gls{iid} channel realizations.}
	\label{fig:benchNC}
\end{figure}
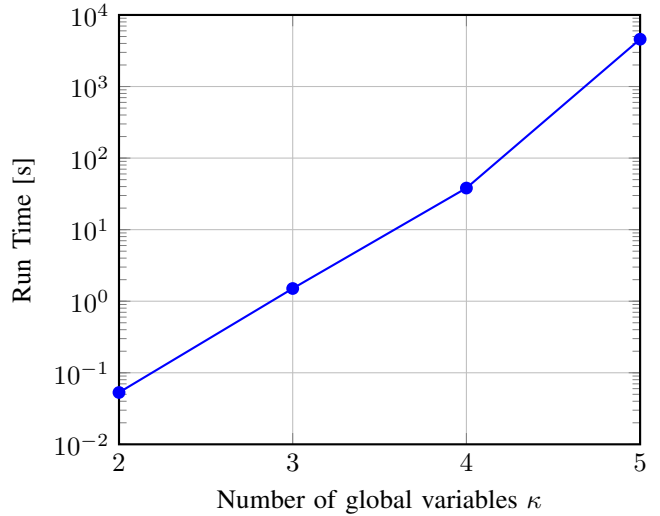

Finally, \cref{fig:benchNC} verifies the exponential complexity in the number of global variables. With the same paramters as in \cref{fig:benchUE}, 
\cref{fig:benchNC} displays the run time for a fixed number of $K=7$ users over the number of global variables $\kappa$. Due to the logarithmical y-axis, exponential growth corresponds to (at least) a linear function in \cref{fig:benchNC}.

\begin{figure}
	\centering
	\tikzexternaldisable
	\tikzsetnextfilename{benchUE}
	\tikzpicturedependsonfile{benchUE.dat}
	\begin{tikzpicture}
		\begin{axis} [
				thick,
				xlabel={Number of Users $K$},
				ylabel={Run Time [s]},
				ylabel near ticks,
				xlabel near ticks,
				grid,
				ymax = 1,
				ymin = 1e-2,
				xmax = 50,
				xmin = 10,
				minor x tick num = 4,
				xtick distance = 5,
				no markers,
				legend pos=north west,
				legend cell align=left,
				cycle list name=default,
			]

			\pgfplotstableread[col sep=comma]{benchUE.dat}\tbl

			\addplot table[y=NC2, restrict x to domain=10:50] {\tbl};
			\addlegendentry{$\kappa = 2$ global variables};

			\addplot table[y=NC3, restrict x to domain=10:50] {\tbl};
			\addlegendentry{$\kappa = 3$ global variables};
		\end{axis}
	\end{tikzpicture}
	\vspace{-2ex}
	\caption{Average run time of \cref{alg:sitbb} to solve \cref{opt:gic} as a function of the total number of users for fixed number of $\kappa$ global variables. The abcissa is proportional to the number of non-global variables ($K- \kappa$). Averaged over 1,000 \gls{iid} channel realizations.}
	\label{fig:benchUE}
\end{figure}
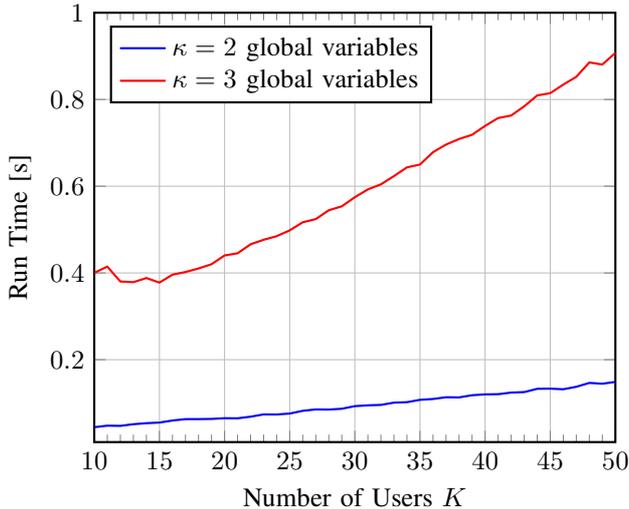

\subsection{Implementation Details}
\Cref{alg:sitbb} was implemented in C++ and compiled with GCC~7.3. Linear optimization problems are solved with Gurobi~8 \cite{gurobi8} and non-linear problems with Mosek~8.1 \cite{mosek} using the kindly provided academic licenses. The complete source code is available on GitHub \cite{github}.

All computations were done on TU Dresden's Bull HPC-Cluster Taurus. Reported performance results were obtained on Intel Haswell nodes with Xeon E5-2680 v3 CPUs running at \unit[2.50]{GHz}.
We thank the Center for Information Services and High Performance Computing (ZIH) at TU Dresden for generous allocations of computer time.

\section{Conclusions} \label{sec:conclusions}
We established $\varepsilon$-essential feasibility as an important concept towards numerical stable global optimization algorithms and introduced the accompanying \acrfull{sit} approach. Based on these concepts we built a novel global optimization framework tailored to resource allocation problems that preserves the computational complexity in the number of non-global variables, inherently supports fractional objectives, avoids numerical problems with non-robust feasible sets, and is four orders of magnitude faster than state-of-the-art algorithms.
We applied the proposed scheme to a resource allocation problem for the Gaussian \cgls{mwrc} with \cgls{af} relaying.
Reproducible research and easy adoption of this algorithm is enabled by releasing the complete source code on GitHub \cite{github}.

\appendix

\subsection{Proof of \cref{prop:robust}}
	First consider \cref{constrA}. The level set $\mathcal F = \{ \gamma f^-(\vec x, \vec\xi) - f^+(\vec x, \vec\xi) \le 0 \}$ is closed and convex \cite[Thms.~4.6\,\&\,7.1]{Rockafellar1970}. Hence, $\mathcal D = \mathcal C \cap \mathcal F$ is also closed and convex. Every nonempty closed convex set is robust \cite[Thm.~6.3]{Rockafellar1970}. Thus, $\mathcal D$ is robust or \cref{opt:dualRA} is infeasible.

	Now for \cref{constrB}. The feasible set of \cref{opt:dualRA} is $\mathcal D = \{ \vec x\in\mathcal X, \vec\xi\in\Xi : f_\xi(\vec \xi) + f_x(\vec x) \le 0 \}$ with $f_\xi(\vec\xi) = \gamma f^-_\xi(\vec\xi) - f^+_\xi(\vec\xi)$ and $f_x(\vec x) = \gamma f^-_x(\vec x) - f^+_x(\vec x)$. By assumption, $f_\xi(\vec\xi)$ is an \cgls{lsc} convex and $f_x(\vec x)$ an \cgls{lsc} increasing (decreasing) function. Further, $\mathcal X$ is normal (conormal) within a box and $\Xi$ is convex. Observe that $\mathcal D$ is a convex set in $\vec\xi$ since for fixed $\vec x$, $f_x(\vec x)$ is a constant, $\tilde f_\xi(\vec \xi) = f_\xi(\vec\xi) + \mathrm{const.}$ is a convex function, and $\{\vec\xi : \tilde f_\xi(\vec \xi) \le 0\}$ is a closed convex set \cite[Thms.~4.6\,\&\,7.1]{Rockafellar1970}. By the same argument, $\tilde f_x(\vec x)$ is an increasing (decreasing) function and $\{\vec x : \tilde f_x(\vec x) \le 0\}$ is a closed normal (conormal) set \cite[Prop.~11.2]{Tuy2016}. Thus, $\mathcal D$ is normal (conormal) in a box in $\vec x$. Neither closed convex nor closed (co-)normal sets have any isolated feasible points \cite{Tuy2009}. Since $\mathcal D$ is either convex or (co-)normal in each coordinate the proposition is proven. \hfill\IEEEQED

\subsection{Proof of \cref{prop:underestimator}}
For all $\vec x \in \mathcal M$,
$g_i^-(\bar{\vec x}^\ast) \ge g_i^-(\vec x)$ and hence also $g_i^+(\vec x, \vec\xi) - g_i^-(\bar{\vec x}^\ast) \le g_i^+(\vec x, \vec\xi) - g_i^-(\vec x)$ with equality at $\vec x = \bar{\vec x}^\ast$.
It remains to show that for all real-valued functions $h_1, h_2, \dots, \ubar h_1, \ubar h_2, \dots$ satisfying $h_i \ge \ubar h_i$ and $h_i(\vec y) = \ubar h_i(\vec y)$ for all $i$ and some point $\vec y$, $\max_i\{h_i\} \ge \max_i\{\ubar h_i\}$ and $\max_i\{h_i(\vec y)\} = \max_i\{\ubar h_i(\vec y)\}$ holds.

Consider the case with two functions and assume that $\max\{h_1, h_2\} < \max\{\ubar h_1, \ubar h_2\}$. Since $h_i \ge \ubar h_i$, this can only hold if $\max\{h_1, h_2\} = h_1$ and $\max\{\ubar h_1, \ubar h_2\} = \ubar h_2$ or vice versa. This implies $h_1 \ge h_2 \ge \ubar h_2$ which contradicts the assumption. The generalization to arbitrarily many functions follows by induction.
Finally, if $\ubar h_i(\vec y) = h_i(\vec y)$ for all $i$, then $\min_i\{\ubar h_i(\vec y)\} = \min_i\{h_i(\vec y)\}$. \hfill\IEEEQED

\subsection{Proof of \cref{thm:convergence}}
Convergence of the \cgls{bb} procedure is mostly established by \cref{prop:bb} and \cite[Prop.~6.2]{Tuy2016}.
	It remains to show that choosing $\beta(\mathcal M_i)$ as the optimal value of \cref{opt:bound} or \cref{opt:bound3} satisfies \cref{eq:bbconv} and that continuing the \cgls{bb} procedure after updating $\gamma$ preserves convergence.

	First, observe that neither \cref{opt:bound} nor \cref{opt:bound3} relax the feasible set $\mathcal D$ of \cref{opt:dualRA}. Hence, the first condition in \cref{eq:bbconv} is always satisfied. Further,
\begin{align*} 
	\beta(\mathcal M^k) &= \underset{\vec\xi\in\mathcal D_{\vec x^k}}{\text{min}} \underset{i=1, 2, \dots, m}{\text{max}} \left\{ g_i^+(\vec x^k, \vec\xi) - g_i^-(\bar{\vec x}^\ast_{\mathcal M_i}) \right\}
\intertext{and}
	\underset{\vec\xi\in\mathcal D_{\vec v^k}}{\text{min}} g(\vec v^k, \vec\xi) &= \underset{\vec\xi\in\mathcal D_{\vec v^k}}{\text{min}}
	\underset{i=1, 2, \dots, m}{\text{max}} \left( g_i^+(\bar{\vec x}^\ast_{\mathcal M^k}, \vec\xi) - g_i^-(\bar{\vec x}^\ast_{\mathcal M^k}) \right)
\end{align*}
Thus, $\beta(\mathcal M^k) \rightarrow \operatorname{\text{min}}_{\vec\xi\in\mathcal D_{\vec v^k}} g(\vec v^k, \vec\xi)$ as $\norm{\vec x^k - \vec v^k} = \norm{\vec x^k - \bar{\vec x}^\ast_{\mathcal M^k}} \rightarrow 0$ and \cref{eq:bbconv} is satisfied.

Next, let $\{ \gamma_k \}$ be the sequence of updated gammas, and observe that this sequence is increasing, i.e., $\gamma_{k+1} \ge \gamma_k$. Thus, the feasible sets of \cref{opt:bound} form a decreasing sequence of sets
	\begin{equation*}
		\bigg\{ \vec x \in\tilde{\mathcal C} \,\bigg|\, \frac{f^+(\vec x, \vec\xi)}{f^-(\vec x, \vec\xi)}\ge\gamma_{k+1} \bigg\}
		\subseteq
		\bigg\{ \vec x \in\tilde{\mathcal C} \,\bigg|\, \frac{f^+(\vec x, \vec\xi)}{f^-(\vec x, \vec\xi)}\ge\gamma_k \bigg\}
	\end{equation*}
	with $\tilde{\mathcal C} = \{ (\vec x, \vec\xi) \,|\, (\vec x, \vec\xi)\in\mathcal C,\ \vec x\in\mathcal M \}$. Therefore, the optimal value of the bound \cref{opt:bound} is increasing with $k$
	\begin{equation}
		\label{thm:convergence:1}
		\begin{aligned}
			&\min_{\vec x, \vec\xi} \bigg\{ \tilde g(\vec x, \vec\xi) \,\bigg|\, \frac{f^+(\vec x, \vec\xi)}{f^-(\vec x, \vec\xi)}\ge\gamma_{k+1},\ (\vec x, \vec\xi)\in\tilde{\mathcal C} \bigg\} \\
			\ge
			&\min_{\vec x, \vec\xi} \bigg\{ \tilde g(\vec x, \vec\xi) \,\bigg|\, \frac{f^+(\vec x, \vec\xi)}{f^-(\vec x, \vec\xi)}\ge\gamma_{k},\ (\vec x, \vec\xi)\in\tilde{\mathcal C} \bigg\}
		\end{aligned}
	\end{equation}
	where $\tilde g(\vec x, \vec\xi) = \max_{i=1, 2, \dots, m} \left\{ g_i^+(\vec x, \vec\xi) - g_i^-(\bar{\vec x}^\ast_{\mathcal M_i}) \right\}$.
	Hence, every box eliminated due to the deletion criterion $\beta(\mathcal M) > -\varepsilon$ in a \cgls{bb} procedure with $\gamma_k$, would also be eliminated in a procedure with $\gamma_{k+1}$. It follows that the set holding the boxes $\mathscr R_k$ remains valid after updating $\gamma$. In particular, no box is eliminated prematurely. Thus, restarting the \cgls{bb} procedure after updating $\gamma$ is not necessary.
Convergence of \cref{alg:sitbb} is finite since the underlying \cgls{bb} procedure is finite and $\{ \gamma_k \}$ is bounded. The same argument can be made for \cref{opt:bound3}.
\hfill\IEEEQED

\subsection{Proof sketch for \cref{lem:HK}}
Extend \cite[Thm.~1]{Matthiesen2015a} to Gaussian channels using the standard procedure in \cite[Sect.~3.4.1]{ElGamal2011}. 
Evaluate it with Gaussian inputs $U_k\sim\mathcal{CN}(0, P_k^c)$ and $X_k = U_k + V_k$ with $V_k\sim\mathcal{CN}(0, P_k^p)$,
and $\mathds E[X_0^2] = P_0$ to obtain the rate expressions above with $\widetilde{g}_k = \abs{g_k}^2 \frac{P_0}{N_k}$.
The achievable rates are increasing in $P_0$. Thus, $P_0 = \bar P_0$ is rate-optimal. \hfill\IEEEQED

\subsection{Proof of \cref{prop:equiv}}
Let $\mathcal F$ be the feasible set of \cref{opt:HK} without the first constraint. Then, we have
\begin{align*}
	&\max_{(\vec S, \vec R) \in \mathcal F} \{ f(\vec S, \vec R) \,|\, \forall i: \vec a_i^T \vec R \le \sum_j \Capa\Big( \frac{\vec b_{i,j}^T \vec S}{\gamma_{\kappa(i,j)}(\vec S)} \Big) \Big\} \\
	=&\max_{\substack{(\vec S, \vec R) \in \mathcal F\\\mathclap{y = \sum_{k\in\mathcal K} \abs{h_k}^2 S^c_k}}} \{ f(\vec S, \vec R) \,|\, \forall i: \vec a_i^T \vec R \le \sum_j \Capa\Big( \frac{\vec b_{i,j}^T \vec S}{\gamma_{\kappa(i,j)}(\vec S^p, y)} \Big) \Big\} \\
	\le&\max_{\substack{(\vec S, \vec R) \in \mathcal F\\\mathclap{y \ge \sum_{k\in\mathcal K} \abs{h_k}^2 S^c_k}}} \{ f(\vec S, \vec R) \,|\, \forall i: \vec a_i^T \vec R \le \sum_j \Capa\Big( \frac{\vec b_{i,j}^T \vec S}{\gamma_{\kappa(i,j)}(\vec S^p, y)} \Big) \Big\},
\end{align*}
since relaxing a constraint does not decrease the optimal value.
Conversely, $\gamma_k(\vec S^p, y)$ is increasing in $y$ and, thus, the \cglspl{rhs} of the constraints are decreasing in $y$. Since $f(\vec S, \vec R)$ and $\vec a_i^T \vec R$ are increasing in $\vec R$, the \cgls{rhs} should be as large as possible. Thus, the optimal $y$ is as small as possible and
\begin{align*}
	&\max_{\substack{(\vec S, \vec R) \in \mathcal F\\\mathclap{y \ge \sum_{k\in\mathcal K} \abs{h_k}^2 S^c_k}}} \{ f(\vec S, \vec R) \,|\, \forall i: \vec a_i^T \vec R \le \sum_j \Capa\Big( \frac{\vec b_{i,j}^T \vec S}{\gamma_{\kappa(i,j)}(\vec S^p, y)} \Big) \Big\} \\
	\le& \max_{\substack{(\vec S, \vec R) \in \mathcal F\\\mathclap{y = \sum_{k\in\mathcal K} \abs{h_k}^2 S^c_k}}} \{ f(\vec S, \vec R) \,|\, \forall i: \vec a_i^T \vec R \le \sum_j \Capa\Big( \frac{\vec b_{i,j}^T \vec S}{\gamma_{\kappa(i,j)}(\vec S^p, y)} \Big) \Big\}.
\end{align*}

This establishes \cref{prop:equiv}. \hfill\IEEEQED

\balance
\bibliography{IEEEabrv,paper.bib}
\end{document}

%% file: acronyms.tex
\makeglossaries

\newacronym{dsl}{DSL}{digital subscriber line}
\newacronym{wsee}{WSEE}{weighted sum energy efficiency}
\newacronym{mmwave}{mmWave}{millimeter wave}
\newacronym{dfg}{DFG}{Deutsche Forschungsgemeinschaft}
\newacronym{haec}{HAEC}{Highly Adaptive Energy-Efficient Computing}
\newacronym{hpc}{HPC}{High Performance Computing}
\newacronym{mac}{MAC}{multiple-access channel}
\newacronym{bc}{BC}{broadcast channel}
\newacronym{siso}{SISO}{single-input single-output}
\newacronym{simo}{SIMO}{single-input multiple-output}
\newacronym{miso}{MISO}{multiple-input single-output}
\newacronym{mimo}{MIMO}{multiple-input multiple-output}
\newacronym{af}{AF}{amplify-and-forward}
\newacronym{df}{DF}{decode-and-forward}
\newacronym{cf}{CF}{compress-and-forward}
\newacronym{mwrc}{MWRC}{multi-way relay channel}
\newacronym{dmmwrc}{DM-MWRC}{discrete memoryless multi-way relay channel}
\newacronym{pde}{PDE}{partial data exchange}
\newacronym{fde}{FDE}{full data exchange}
\newacronym{iid}{i.i.d.\@}{independent and identically distributed}
\newacronym{di}{DI} {difference of increasing}
\newacronym{dc}{DC}{difference of convex}
\newacronym{mm}{MM}{mixed monotonic}
\newacronym{mmp}{MMP}{mixed monotonic programming}
\newacronym{awgn}{AWGN}{additive white Gaussian noise}
\newacronym{awg}{AWG}{additive white Gaussian}
\newacronym{sic}{SIC}{successive interference cancellation}
\newacronym{snr}{SNR}{signal-to-noise ratio}
\newacronym{sinr}{SINR}{signal to interference plus noise ratio}
\newacronym{inr}{INR}{interference to noise ratio}
\newacronym{zf}{ZF}{zero-forcing}
\newacronym{mrt}{MRT}{maximum ratio transmission}
\newacronym{mmse}{MMSE}{minimum mean square error}
\newacronym{sud}{SUD}{single user decoding}
\newacronym{dof}{DoF}{degrees of freedom}
\newacronym{gdof}{GDoF}{generalized degrees of freedom}
\newacronym{nnc}{NNC}{noisy network coding}
\newacronym{dmn}{DMN}{discrete memoryless network}
\newacronym{csi}{CSI}{channel state information}
\newacronym{pmf}{pmf}{probability mass function}
\newacronym{dmic}{DM-IC}{discrete memoryless interference channel}
\newacronym{ic}{IC}{interference channel}
\newacronym{gic}{GIC}{Gaussian interference channel}
\newacronym{if}{IF}{interference}
\newacronym{ee}{EE}{energy efficiency}
\newacronym{gee}{GEE}{global energy efficiency}
\newacronym{tin}{TIN}{treating interference as noise}
\newacronym{snd}{SND}{simultaneous non-unique decoding}
\newacronym{sd}{SD}{simultaneous decoding}
\newacronym{hk}{HK}{Han-Kobayashi}
\newacronym{rs}{RS}{rate splitting}
\newacronym{rf}{RF}{radio frequency}
\newacronym{pa}{PA}{power amplifier}
\newacronym{lna}{LNA}{low noise amplifier}
\newacronym{lo}{LO}{local oscillator}
\newacronym{adc}{ADC}{analog-to-digital converter}
\newacronym{dac}{DAC}{digital-to-analog converter}
\newacronym{dsp}{DSP}{digital signal processing}
\newacronym{brd}{BRD}{best response dynamics}
\newacronym{br}{BR}{best response}
\newacronym{ne}{NE}{Nash equilibrium}
\newacronym{lhs}{LHS}{left-hand side}
\newacronym{rhs}{RHS}{right-hand side}
\newacronym{ran}{RAN}{radio access network}
\newacronym{qos}{QoS}{Quality of Service}
\newacronym{ngmn}{NGMN}{Next Generation Mobile Networks}
\newacronym{cap}{CAP}{Capacity Adaptation}
\newacronym{bwa}{BW}{Bandwidth Adaptation}
\newacronym{prb}{PRB}{physical resource block}
\newacronym{se}{SE}{spectral efficiency}
\newacronym{tp}{TP}{throughput}
\newacronym{bs}{BS}{base station}
\newacronym{mop}{MOP}{multi-objective optimization problem}
\newacronym{gda}{GDA}{generalized Dinkelbach's algorithm}
\newacronym{midcp}{MIDCP}{mixed integer disciplined convex programming}
\newacronym{lp}{LP}{linear program}
\newacronym{brb}{BRB}{branch reduce and bound}
\newacronym{bb}{BB}{branch and bound}
\newacronym{sit}{SIT}{successive incumbent transcending}
\newacronym{oma}{OMA}{orthogonal multiple access}
\newacronym{noma}{NOMA}{non-orthogonal multiple access}
\newacronym{wlog}{w.l.o.g.\@}{without loss of generality}
\newacronym{lsc}{l.s.c.\@}{lower semi-continuous}
\newacronym{usc}{u.s.c.\@}{upper semi-continuous}
\newacronym{kkt}{KKT}{Karush-Kuhn-Tucker}
\newacronym{ptp}{PTP}{point-to-point}

%% file: paper.bbl
\begin{thebibliography}{10}
\providecommand{\url}[1]{#1}
\csname url@samestyle\endcsname
\providecommand{\newblock}{\relax}
\providecommand{\bibinfo}[2]{#2}
\providecommand{\BIBentrySTDinterwordspacing}{\spaceskip=0pt\relax}
\providecommand{\BIBentryALTinterwordstretchfactor}{4}
\providecommand{\BIBentryALTinterwordspacing}{\spaceskip=\fontdimen2\font plus
\BIBentryALTinterwordstretchfactor\fontdimen3\font minus
  \fontdimen4\font\relax}
\providecommand{\BIBforeignlanguage}[2]{{%
\expandafter\ifx\csname l@#1\endcsname\relax
\typeout{** WARNING: IEEEtran.bst: No hyphenation pattern has been}%
\typeout{** loaded for the language `#1'. Using the pattern for}%
\typeout{** the default language instead.}%
\else
\language=\csname l@#1\endcsname
\fi
#2}}
\providecommand{\BIBdecl}{\relax}
\BIBdecl

\bibitem{spawc18}
B.~Matthiesen and E.~A. Jorswieck, ``Optimal resource allocation for
  non-regenerative multiway relaying with rate splitting,'' in \emph{Proc.
  {IEEE} 19th Int. Workshop Signal Process. Adv. Wireless Commun. ({SPAWC})},
  Kalamata, Greece, Jun. 2018.

\bibitem{icassp2019}
------, ``Global energy efficiency maximization in non-orthogonal interference
  networks,'' in \emph{Proc. IEEE Int. Conf. Acoustics, Speech and Signal
  Processing (ICASSP)}, Brighton, United Kingdom, May 2019.

\bibitem{Tuy2016}
H.~Tuy, \emph{Convex Analysis and Global Optimization}, ser. Springer
  Optimization and Its Applications.\hskip 1em plus 0.5em minus 0.4em\relax
  Springer, 2016.

\bibitem{camsap17}
B.~Matthiesen and E.~A. Jorswieck, ``Weighted sum rate maximization for
  non-regenerative multi-way relay channels with multi-user decoding,'' in
  \emph{Proc. {IEEE} 7th Int. Workshop Comput. Adv. Multi-Sensor Adaptive
  Process. ({CAMSAP})}, Cura\c cao, Dutch Antilles, Dec. 2017.

\bibitem{Han2008}
Z.~Han and K.~J.~R. Liu, \emph{Resource Allocation for Wireless Networks:
  Basics, Techniques, and Applications}.\hskip 1em plus 0.5em minus 0.4em\relax
  Cambridge University Press, 2008.

\bibitem{Horst1996}
R.~Horst and H.~Tuy, \emph{Global Optimization: Deterministic Approaches},
  3rd~ed.\hskip 1em plus 0.5em minus 0.4em\relax Springer, 1996.

\bibitem{bandemer2015}
B.~Bandemer, A.~El~Gamal, and Y.-H. Kim, ``Optimal achievable rates for
  interference networks with random codes,'' \emph{{IEEE} Trans. Inf. Theory},
  vol.~61, no.~12, pp. 6536--6549, Oct. 2015.

\bibitem{Khachiyan1979}
L.~G. Khachiyan, ``A polynomial algorithm in linear programming,''
  \emph{Doklady Academii Nauk SSSR}, vol. 244, pp. 1093--1096, 1979.

\bibitem{Karmarkar1984}
N.~Karmarkar, ``A new polynomial-time algorithm for linear programming,''
  \emph{Combinatorica}, vol.~4, no.~4, pp. 373--395, Dec. 1984.

\bibitem{Ben-Tal2009}
A.~Ben-Tal, L.~El~Ghaoui, and A.~Nemirovski, \emph{Robust optimization}.\hskip
  1em plus 0.5em minus 0.4em\relax Princeton University Press, 2009.

\bibitem{Vorobyov2003}
S.~A. Vorobyov, A.~B. Gershman, and Z.-Q. Luo, ``Robust adaptive beamforming
  using worst-case performance optimization: A solution to the signal mismatch
  problem,'' \emph{{IEEE} Trans. Signal Process.}, vol.~51, no.~2, pp.
  313--324, Feb. 2003.

\bibitem{Bjoernson2012}
E.~Bj\"{o}rnson, G.~Zheng, M.~Bengtsson, and B.~Ottersten, ``Robust monotonic
  optimization framework for multicell {MISO} systems,'' \emph{{IEEE} Trans.
  Signal Process.}, vol.~60, no.~5, pp. 2508--2523, Jan. 2012.

\bibitem{Tuy2005a}
H.~Tuy, ``Robust solution of nonconvex global optimization problems,'' \emph{J.
  Global Optim.}, vol.~32, no.~2, pp. 307--323, Jun. 2005.

\bibitem{Tuy2009}
------, ``{$\mathcal{D(C)}$}-optimization and robust global optimization,''
  \emph{J. Global Optim.}, vol.~47, no.~3, pp. 485--501, Oct. 2009.

\bibitem{Zappone2017}
A.~Zappone, E.~Bj\"{o}rnson, L.~Sanguinetti, and E.~A. Jorswieck, ``Globally
  optimal energy-efficient power control and receiver design in wireless
  networks,'' \emph{{IEEE} Trans. Signal Process.}, vol.~65, no.~11, pp.
  2844--2859, Jun. 2017.

\bibitem{Phan2012}
A.~H. Phan, H.~D. Tuan, and H.~H. Kha, ``{D.C.} iterations for {SINR} maximin
  multicasting in cognitive radio,'' in \emph{Proc. 6th Int. Conf. Signal
  Process. Commun. Syst. ({ICSPCS})}, Gold Coast, Australia, Dec. 2012.

\bibitem{Xu2010}
J.~Xu, Z.~Miao, and Q.~Liu, ``New method to get essential efficient solution
  for a class of {D.C.} multiobjective problem,'' in \emph{Proc. Int. Conf.
  Comput., Mechatronics, Control, Electron. Eng. ({CMCE})}, Changchun, China,
  Oct. 2010.

\bibitem{Palomar2006}
D.~P. Palomar and M.~Chiang, ``A tutorial on decomposition methods for network
  utility maximization,'' \emph{{IEEE} J. Sel. Areas Commun.}, vol.~24, no.~8,
  pp. 1439--1451, Aug. 2006.

\bibitem{Palomar2005}
D.~P. Palomar, ``Convex primal decomposition for multicarrier linear {MIMO}
  transceivers,'' \emph{{IEEE} Trans. Signal Process.}, vol.~53, no.~12, pp.
  4661--4674, Dec. 2005.

\bibitem{Kaleva2016}
J.~Kaleva, A.~T\"{o}lli, and M.~Juntti, ``Decentralized sum rate maximization
  with {QoS} constraints for interfering broadcast channel via successive
  convex approximation,'' \emph{{IEEE} Trans. Signal Process.}, vol.~64,
  no.~11, pp. 2788--2802, Jun. 2016.

\bibitem{Tervo2018}
O.~Tervo, H.~Pennanen, D.~Christopoulos, S.~Chatzinotas, and B.~Ottersten,
  ``Distributed optimization for coordinated beamforming in multicell
  multigroup multicast systems: Power minimization and {SINR} balancing,''
  \emph{{IEEE} Trans. Signal Process.}, vol.~66, no.~1, pp. 171--185, Jan.
  2018.

\bibitem{Rossi2011}
M.~Rossi, A.~M. Tulino, O.~Simeone, and A.~M. Haimovich, ``Non-convex utility
  maximization in {Gaussian} {MISO} broadcast and interference channels,'' in
  \emph{Proc. {IEEE} Int. Conf. Acoust., Speech, Signal Process. (ICASSP)},
  Prague, Czech Republic, May 2011.

\bibitem{Tuy2000}
H.~Tuy, ``Monotonic optimization: Problems and solution approaches,''
  \emph{SIAM J. Optimization}, vol.~11, no.~2, pp. 464--494, Feb. 2000.

\bibitem{Qian2009}
L.~P. Qian, Y.~J. Zhang, and J.~Huang, ``{MAPEL}: Achieving global optimality
  for a non-convex wireless power control problem,'' \emph{{IEEE} Trans.
  Wireless Commun.}, vol.~8, no.~3, pp. 1553--1563, 2009.

\bibitem{Bjornson2013}
E.~Bj\"{o}rnson and E.~A. Jorswieck, \emph{Optimal Resource Allocation in
  Coordinated Multi-Cell Systems}, ser. Found. Trends Commun. Inf.
  Theory.\hskip 1em plus 0.5em minus 0.4em\relax Now Publishers, 2013, vol.~9,
  no. 2-3.

\bibitem{Zappone2016}
A.~Zappone, L.~Sanguinetti, G.~Bacci, E.~A. Jorswieck, and M.~Debbah,
  ``Energy-efficient power control: A look at {5G} wireless technologies,''
  \emph{{IEEE} Trans. Signal Process.}, vol.~64, no.~7, pp. 1668--1683, Apr.
  2016.

\bibitem{Tervo2017}
O.~Tervo, A.~T\"olli, M.~Juntti, and L.-N. Tran, ``Energy-efficient beam
  coordination strategies with rate dependent processing power,'' \emph{{IEEE}
  Trans. Signal Process.}, vol.~65, no.~22, pp. 6097--6112, Nov. 2017.

\bibitem{Ng2013}
D.~W.~K. Ng, E.~S. Lo, and R.~Schober, ``Wireless information and power
  transfer: energy efficiency optimization in {OFDMA} systems,'' \emph{{IEEE}
  Trans. Wireless Commun.}, vol.~12, no.~12, pp. 6352--6370, Dec. 2013.

\bibitem{Ng2012}
------, ``Energy-efficient resource allocation in multi-cell {OFDMA} systems
  with limited backhaul capacity,'' \emph{{IEEE} Trans. Wireless Commun.},
  vol.~11, no.~10, pp. 3618--3631, Oct. 2012.

\bibitem{github}
\BIBentryALTinterwordspacing
B.~Matthiesen. (2018) Accompanying source code. [Online]. Available:
  \url{https://github.com/bmatthiesen/efficient-global-opt}
\BIBentrySTDinterwordspacing

\bibitem{Nesterov1994}
Y.~Nesterov and A.~Nemirovskii, \emph{Interior-Point Polynomial Algorithms in
  Convex Programming}, ser. SIAM Studies in Applied Mathematics.\hskip 1em plus
  0.5em minus 0.4em\relax SIAM, 1994.

\bibitem{Ben-Tal2001}
A.~Ben-Tal and A.~Nemirovskii, \emph{Lectures on Modern Convex Optimization:
  Analysis, Algorithms, and Engineering Applications}, ser. MPS-SIAM Series on
  Optimization.\hskip 1em plus 0.5em minus 0.4em\relax SIAM, 2001.

\bibitem{Nemirovskii1983}
A.~S. Nemirovskii and D.~B. Yudin, \emph{Problem complexity and method
  efficiency in optimization.}\hskip 1em plus 0.5em minus 0.4em\relax Wiley,
  1983.

\bibitem{wcnc18}
B.~Matthiesen, Y.~Yang, and E.~A. Jorswieck, ``Optimization of weighted
  individual energy efficiencies in interference networks,'' in \emph{Proc.
  {IEEE} Wireless Commun. Netw. Conf. ({WCNC})}, Barcelona, Spain, Apr. 2018.

\bibitem{Haykin2005}
S.~Haykin, ``Cognitive radio: brain-empowered wireless communications,''
  \emph{{IEEE} J. Sel. Areas Commun.}, vol.~23, no.~2, pp. 201--220, Feb. 2005.

\bibitem{Yu2002}
\BIBentryALTinterwordspacing
W.~Yu, G.~Ginis, and J.~Cioffi, ``Distributed multiuser power control for
  digital subscriber lines,'' \emph{{IEEE} J. Sel. Areas Commun.}, vol.~20,
  no.~5, pp. 1105--1115, Jun. 2002. [Online]. Available:
  \url{http://dx.doi.org/10.1109/JSAC.2002.1007390}
\BIBentrySTDinterwordspacing

\bibitem{Zadeh1963}
L.~A. Zadeh, ``Optimality and non-scalar-valued performance criteria,''
  \emph{{IEEE} Trans. Autom. Control}, vol.~8, no.~1, pp. 59--60, Jan. 1963.

\bibitem{Jorswieck2002}
E.~A. Jorswieck and H.~Boche, ``Rate balancing for the multi-antenna {G}aussian
  broadcast channel,'' in \emph{Proc. {IEEE} 7th Int. Symp. Spread Spectr.
  Techn. Appl.}, Prague, Czech Republic, Sep. 2002.

\bibitem{Zhang2010a}
R.~Zhang and S.~Cui, ``Cooperative interference management with {MISO}
  beamforming,'' \emph{{IEEE} Trans. Signal Process.}, vol.~58, no.~10, pp.
  5450--5458, Oct. 2010.

\bibitem{Tassiulas1992}
L.~Tassiulas and A.~Ephremides, ``Stability properties of constrained queueing
  systems and scheduling policies for maximum throughput in multihop radio
  networks,'' \emph{{IEEE} Trans. Autom. Control}, vol.~37, no.~12, pp.
  1936--1948, Dec. 1992.

\bibitem{Neely2003}
M.~J. Neely, E.~Modiano, and C.~E. Rohrs, ``Power allocation and routing in
  multibeam satellites with time-varying channels,'' \emph{{IEEE/ACM} Trans.
  Netw.}, vol.~11, no.~1, pp. 138--152, Feb. 2003.

\bibitem{Tse1998}
D.~N.~C. Tse and S.~V. Hanly, ``Multiaccess fading channels. {I.} {P}olymatroid
  structure, optimal resource allocation and throughput capacities,''
  \emph{{IEEE} Trans. Inf. Theory}, vol.~44, no.~7, pp. 2796--2815, 1998.

\bibitem{Jindal2005}
N.~Jindal, W.~Rhee, S.~Vishwanath, S.~A. Jafar, and A.~Goldsmith, ``Sum power
  iterative water-filling for multi-antenna {G}aussian broadcast channels,''
  \emph{{IEEE} Trans. Inf. Theory}, vol.~51, no.~4, pp. 1570--1580, Apr. 2005.

\bibitem{Jorswieck2007a}
E.~A. Jorswieck and H.~Boche, ``On the performance optimization in multiuser
  {MIMO} systems,'' \emph{Trans. Emerg. Telecommun. Technol.}, vol.~18, no.~3,
  pp. 287--304, Apr. 2007.

\bibitem{Jorswieck2010a}
E.~A. Jorswieck and E.~G. Larsson, ``Monotonic optimization framework for the
  two-user {MISO} interference channel,'' \emph{{IEEE} Trans. Commun.},
  vol.~58, no.~7, pp. 2159--2168, Jul. 2010.

\bibitem{Isheden2012}
C.~Isheden, Z.~Chong, E.~A. Jorswieck, and G.~Fettweis, ``Framework for
  link-level energy efficiency optimization with informed transmitter,''
  \emph{{IEEE} Trans. Wireless Commun.}, vol.~11, no.~8, pp. 2946--2957, Aug.
  2012.

\bibitem{Charnes1962}
A.~Charnes and W.~W. Cooper, ``Programming with linear fractional
  functionals,'' \emph{Naval Res. Logistics Quart.}, vol.~9, no. 3-4, pp.
  181--186, 1962.

\bibitem{dinkelbach1967}
W.~Dinkelbach, ``On nonlinear fractional programming,'' \emph{Manage. Sci.},
  vol.~13, no.~7, pp. 492--498, Mar. 1967.

\bibitem{Schaible1983}
S.~Schaible and T.~Ibaraki, ``Fractional programming,'' \emph{Eur. J.
  Operational Res.}, vol.~12, no.~4, pp. 325--338, Apr. 1983.

\bibitem{Schaible1993}
S.~Schaible, ``Fractional programming,'' in \emph{Handbook of global
  optimization}, R.~Horst and P.~M. Pardalos, Eds.\hskip 1em plus 0.5em minus
  0.4em\relax Kluwer Academic Publishers, 1993.

\bibitem{Zappone2015}
A.~Zappone and E.~A. Jorswieck, \emph{Energy Efficiency in Wireless Networks
  via Fractional Programming Theory}, ser. Found. Trends Commun. Inf.
  Theory.\hskip 1em plus 0.5em minus 0.4em\relax Now Publishers, 2015, vol.~11,
  no. 3-4.

\bibitem{Freund2001}
R.~W. Freund and F.~Jarre, ``Solving the sum-of-ratios problem by an
  interior-point method,'' \emph{J. Global Optim.}, vol.~19, no.~1, pp.
  83--102, 2001.

\bibitem{Schaible2003}
S.~Schaible and J.~Shi, ``Fractional programming: The sum-of-ratios case,''
  \emph{Optim. Methods, Softw.}, vol.~18, no.~2, pp. 219--229, 2003.

\bibitem{Crouzeix1985}
J.-P. Crouzeix, J.~A. Ferland, and S.~Schaible, ``An algorithm for generalized
  fractional programs,'' \emph{J. Optim. Theory Appl.}, vol.~47, no.~1, pp.
  35--49, 1985.

\bibitem{Kelly1998}
F.~P. Kelly, A.~K. Maulloo, and D.~K.~H. Tan, ``Rate control for communication
  networks: shadow prices, proportional fairness and stability,'' \emph{J.
  Operational Res. Soc.}, vol.~49, no.~3, pp. 237--252, Mar. 1998.

\bibitem{Barros2006}
J.~Barros and M.~R.~D. Rodrigues, ``Secrecy capacity of wireless channels,'' in
  \emph{Proc. {IEEE} Int. Symp. Inf. Theory ({ISIT})}, Seattle, WA, Jul. 2006.

\bibitem{Knuth1997vol3}
D.~E. Knuth, \emph{The Art of Computer Programming, Volume 3: Sorting and
  Searching}, 2nd~ed.\hskip 1em plus 0.5em minus 0.4em\relax Addison-Wesley,
  1997.

\bibitem{Avriel1988}
M.~Avriel, W.~E. Diewert, S.~Schaible, and I.~Zang, \emph{Generalized
  Concavity}.\hskip 1em plus 0.5em minus 0.4em\relax Plenum Press, 1988.

\bibitem{Gunduz2013}
D.~G\"und\"uz, A.~Yener, A.~Goldsmith, and H.~V. Poor, ``The multiway relay
  channel,'' \emph{{IEEE} Trans. Inf. Theory}, vol.~59, no.~1, pp. 51--63, Jan.
  2013.

\bibitem{Chaaban2015}
A.~Chaaban and A.~Sezgin, \emph{Multi-way Communications: An Information
  Theoretic Perspective}, ser. Found. Trends Commun. Inf. Theory.\hskip 1em
  plus 0.5em minus 0.4em\relax Now Publishers, 2015, vol.~12, no. 3-4.

\bibitem{Matthiesen2015a}
B.~Matthiesen and E.~A. Jorswieck, ``Instantaneous relaying for the 3-way relay
  channel with circular message exchanges,'' in \emph{Proc. 49th Asilomar Conf.
  Signals, Syst., Comput.}, Pacific Grove, CA, Nov. 2015, pp. 475--479.

\bibitem{Han1981}
T.~Han and K.~Kobayashi, ``A new achievable rate region for the interference
  channel,'' \emph{{IEEE} Trans. Inf. Theory}, vol.~27, no.~1, pp. 49--60, Jan.
  1981.

\bibitem{gurobi8}
\BIBentryALTinterwordspacing
{Gurobi Optimization}. (2018) Gurobi optimizer 8.0.1. [Online]. Available:
  \url{http://gurobi.com}
\BIBentrySTDinterwordspacing

\bibitem{Matthiesen2015}
B.~Matthiesen, A.~Zappone, and E.~A. Jorswieck, ``Resource allocation for
  energy-efficient 3-way relay channels,'' \emph{{IEEE} Trans. Wireless
  Commun.}, vol.~14, no.~8, pp. 4454--4468, Aug. 2015.

\bibitem{Bjornson2017}
E.~Bj{\"o}rnson, J.~Hoydis, and L.~Sanguinetti, \emph{Massive {MIMO} networks:
  Spectral, energy, and hardware efficiency}, ser. Found. Trends Signal
  Process.\hskip 1em plus 0.5em minus 0.4em\relax Now Publishers, 2017,
  vol.~11, no. 3-4.

\bibitem{mosek}
\BIBentryALTinterwordspacing
{MOSEK ApS}. (2017) {MOSEK} optimizer 8.1.0.34. [Online]. Available:
  \url{http://mosek.com}
\BIBentrySTDinterwordspacing

\bibitem{Rockafellar1970}
R.~T. Rockafellar, \emph{Convex Analysis}.\hskip 1em plus 0.5em minus
  0.4em\relax Princeton University Press, 1970.

\bibitem{ElGamal2011}
A.~El~Gamal and Y.-H. Kim, \emph{Network Information Theory}.\hskip 1em plus
  0.5em minus 0.4em\relax Cambridge University Press, 2011.

\end{thebibliography}
